\documentclass[useAMS,usegraphicx,usenatbib]{mn2e}
\usepackage{enumerate}
\usepackage{amssymb}
\usepackage{amsmath}
\usepackage{array,color}
\usepackage{epsf}
\usepackage{graphics}
\usepackage{dcolumn}
\usepackage{bm}
\newcommand{\be}{\begin{equation}}
\newcommand{\ee}{\end{equation}}
\newcommand{\ba}{\begin{eqnarray}}
\newcommand{\ea}{\end{eqnarray}}
\newcommand{\bi}{\begin{itemize}}
\newcommand{\ei}{\end{itemize}}
\newcommand{\hMpc}{$h\,$Mpc$^{-1}\,$}
\newcommand{\Mpch}{$h^{-1}\,$Mpc }
\newcommand{\Msun}{$h^{-1}\, M_\odot\,$}
\newcommand{\hMpct}{$h^3\,$Mpc$^{-3}$}

\newcommand{\bfi}{\begin{figure}
\epsfxsize=8.1cm
\epsffile}
\newcommand{\efi}{\end{figure}}
\newcommand{\mnras}{MNRAS}
\newcommand{\apj}{ApJ}
\newcommand{\apjl}{ApJ}

\newcommand{\apjs}{ApJS}
\newcommand{\aap}{AAP}
\newcommand{\prd}{PRD}
\newcommand{\jcap}{Journal of Cosmology and Astroparticle Physics}
\newcommand{\physrep}{Physics Reports}
\newcommand{\araa}{Annual Review of Astronomy \& Astrophysics}

\title[Probing the CGM via kSZ effect]{The kinetic Sunyaev-Zel'dovich 
tomography II: probing the circumgalactic medium
}
\author[J. Shao,  T. Fang]
{Jiawei Shao$^{1}$\thanks{email:jwshao@shao.ac.cn}, Taotao Fang $^{2}$\\
$^{1}$Key Laboratory for Research in Galaxies and Cosmology, Shanghai
Astronomical Observatory, Nandan Road 80, Shanghai, 200030, China \\
$^{2}$Department of Astronomy and Institute for Theoretical Physics and 
Astrophysics, Xiamen University, Xiamen, Fujian 361005, China}
\begin{document}
\maketitle
\begin{abstract}
  We propose the use of the kinetic Sunyaev-Zel'dovich (kSZ) effect to probe 
the circumgalactic medium (CGM), with the aid of a spectroscopic survey 
covering the same area of a SZ survey. One can design an optimal estimator of 
the kSZ effect of the CGM with a matched filter, and construct the cross 
correlation between the estimator and the peculiar velocity recovered from the 
galaxy survey, which can be measured by stacking a number of galaxies. We 
investigate two compelling profiles for the CGM, the MB profile \citep{MB2004} 
and the $\beta$ profile, and estimate the detectability against the synergy of 
a fiducial galaxy survey with number density  $10^{-3}h^3\,$ Mpc$^{-3}$ and an 
ACT-like SZ survey. We show that the shape of the filter does not change much 
with redshift for the $\beta$ profile, while there are significant side lobes 
at $z<0.1$ for the MB profile. By stacking $\sim 10^4$ Milky Way-size halos 
around z $\sim 0.5$, one can get $\gtrsim$ 1 $\sigma$ signal to noise (S/N) for 
the both profiles. The S/N increases with decreasing redshift before it reaches 
a maximum ($\sim$ 7.5 at z $\simeq$ 0.15 for the MB profile, $\sim 19$ at 
$z\simeq 0.03$ for the $\beta$ profile). Due to the large beam size, a 
Planck-like CMB survey can marginally detect the kSZ signal by stacking the 
same number of galaxies at $z<0.1$. The search for the CGM in realistic surveys 
will involve dividing the galaxies into subsamples with similar redshift and 
mass of host halos, and scaling the results presented here to obtain the S/N.
\end{abstract}

\section{Introduction}
\label{sec:intro}
Cosmologists have noticed the $\sim$ 50\% shortfall of the cosmic baryons at low 
redshift compared to the baryon content synthesized in the Big Bang for a long 
time \citep{CO1999,Fukugita1998, Fukugita2004, Bregman2007, Shull2012}. The 
cosmic baryon budget has been predicted from a series of cosmic microwave 
background (CMB) surveys \citep{Spergel2003, Komatsu2009, Larson2011}, e.g.  
$\Omega_b=0.048$ from the most recent measurement \citep{Planck2015XIII}.  
At high redshifts, Ly$\alpha$ forest can account for all baryons expected
\citep{Rauch1997, Tytler2004}. However, careful census at low redshift revealed 
that baryons in collapsed objects such as galaxies, galaxy groups and clusters 
only account for about $\sim$ 10\% of the cosmic mean of the baryonic content, 
and $\sim$30\% of the baryons lie in the gas phase revealed by the Ly$\alpha$ 
absorption line. The rest of cosmic baryons are missing. In light of 
cosmological hydrodynamic simulations, majority of the missing baryons are 
believed to reside in a phase called warm hot intergalactic medium (WHIM) with 
temperature between $10^5-10^7$ K \citep{Dave2001, Cen2005}. The WHIM in the 
absorbers of ultraviolet (UV) and X-ray absorption lines signatured by 
Ly$\alpha$, OVI, OVII and other highly ionized metals has been extensively 
investigated \citep[e.g.][]{Tripp2000, Tripp2004, Fang2002, Buote2009, Fang2010,
Tumlinson2011, Tripp2008, Danforth2008, Shull2012}, and the evidence of the
WHIM is accumulating.

A related problem is the galactic missing baryon problem 
\citep{Sommer-Larsen2006, Bregman2007b}. Observationally, the baryonic fraction
increases monotonically with the mass of the gravitational system, and only
galaxy clusters and massive galaxies contain the cosmic mean fraction of
baryons \citep{McGaugh2010}. For low mass galaxies, stars and interstellar
medium (ISM) only account for a small fraction of the baryons, leaving
$\gtrsim$ 80\% of the baryons undetected in observations. The theoretical 
concept of a hot ($\sim$ 10$^6$K) diffuse corona  has been discussed and 
modeled in numerous literatures \citep{Spitzer1956, White1978, WhiteFrenk1991, 
MB2004, Keres2005, Fukugita2006, Henley2013, Henley2015, Faerman2016}. These 
hot coronae could have been shock heated by accretion, or ejected from galaxies 
into the intergalactic medium (IGM) or still detained in the galactic dark halo 
as CGM, depending on the feedback energy in the structure formation process. 
There have been a number of tentative detections of the CGM for both early type 
galaxies and late type galaxies in UV and X-ray band, via both emission and 
absorption \citep[e.g.][]{Anderson2011, Anderson2013, Gupta2012, Thom2012, 
Stocke2013, Dai2012, Owen2009, Werk2014}. On the other hand, for our own 
Galaxy, the presence of hot halo can explain the lack of HI signature in local 
dwarf spheroidal galaxies \citep{Grcevich2009} and the head tail structure of 
the high velocity clouds (HVC)\citep{Putman2011, Gatto2013}. The X-ray emission 
measure also puts a constraint on the baryonic content in the CGM of the Milky 
Way (MW). \citet{Fang2013} discussed that an extended hot gaseous halo profile 
\citep{MB2004} can agree well with the existent constraints. Although no broad 
conclusion of the extent and the profile of the CGM has been drawn, the search 
for the CGM has been a critical issue, and it's possible to detect the CGM 
through a careful design of observation considering the ubiquity of the CGM in 
galaxies.

The Sunyaev-Zel'dovich (SZ) effect \citep{SZ1972} is one of the most promising 
methods of detecting the missing baryons. After the reionization epoch, our 
universe is almost completely ionized after z=6, and free electrons are 
prevailing in galaxy clusters as intracluster medium (ICM) and in between as 
IGM. They will scatter off CMB photons via inverse Compton scattering, and 
generate secondary CMB anisotropies known as the SZ effect. The SZ effect is 
therefore contributed by all electrons, and it can serve as a tool to probe the 
electrons in ionized gas and hence the missing baryons in the ICM/IGM/CGM. Due 
to relatively low density and temperature of the CGM, it's difficult to detect 
them via the thermal SZ (tSZ) effect ($\Delta T \propto \delta^{5/3}$) and 
X-ray emission ($f \propto \delta^2$). The kinetic SZ (kSZ) effect, however, 
directly measures the electron momentum $\delta(1+{\bf v})$ along the line of 
sight, and is thus less weighted toward hot gas (ICM) in galaxy groups and 
clusters. For less massive halos with virial temperature around $10^6$ K, the 
kSZ effect turns out to be larger than the tSZ effect \citep{Birkinshaw1999, 
Carlstrom2002, Singh2015}. Considering the $\sim 50$\% fraction of missing 
baryons, the prevailing IGM/CGM around galaxies can become a detectable kSZ 
effect source.

Blind detection of the kSZ signal from the auto angular correlation alone 
seems impossible because of the overwhelming contamination of cosmic
infrared background (CIB) and the degeneracy between the kSZ effect and the
primary CMB. The state-of-art observations like the Planck satellite
\footnote{http://www.esa.int/Our\_Activities/Space\_Science/Planck}, 
South Pole Telescope\footnote{http://pole.uchicago.edu/} (SPT), and Atacama
Cosmology Telescope \footnote{http://www.princeton.edu/act/} (ACT) alone are
not able to measure the auto-correlation signal of the kinetic SZ effect. Most
of these surveys alone can only put an upper limit of the kSZ effect
\citep{Sievers2013ACT, Reichardt2012}. Taking into account the additional
information of the tSZ bispectrum can help improve the constraint on the kSZ 
effect \citep{Crawford2013}. However, the results inevitably depend on both the 
thermal and kinetic SZ templates.

Since the kSZ effect traces the peculiar momenta of the free electrons and 
thus leaves imprints in the CMB sky, it can be used to constrain cosmological
models or cosmological parameters \citep[e.g.][] { Monteagudo2006,
Bhattacharya2008, Zhang2011, Ma2014, Li2014, Planck2014IntXIII, Zhang2015jcap}.
Recent works have devoted to measuring the kSZ effect on the CMB sky at the
positions of galaxy clusters or galaxies.  Works have focused on measuring the
bulk flow of the local universe using galaxy clusters \citep{Kashlinsky2008,
Kashlinsky2010}, however, the existence of the bulk flow remained controversial
due to low significance \citep{Keisler2009, Mody2012, Feindt2013}. The first
solid detection of the kSZ effect \citep{Hand2012} reported a 3.8 $\sigma$
significance by measuring the pairwise kSZ effect of galaxy clusters, taking 
advantage of the synergy between the BOSS spectroscopic galaxy group catalogue 
and the ACT survey. Alternatively, the kSZ effect was proposed to probe the 
missing baryons \citep{Ho2009, Monteagudo2009, Shao2011b}. Progress has been 
made towards the measurement of the kSZ effect using galaxies. 
\citet{Lavaux2013} claimed the detection of the local bulk flow by measuring 
the kSZ effect toward the direction of nearby galaxies, which indicates the 
existence of hot plasma around galaxies. Moreover, recent works 
\citep{PlanckIntResCV, Monteagudo2015} found the evidence of kSZ signature out
to about three times the mean virial radius of the central galaxy catalogue
(CGC) sample, in which they employed both the pairwise momentum estimator and
the kSZ temperature-velocity field correlation estimator. These results suggest
that the hot plasma around galaxies may be responsible for the missing baryons
of galaxies, and produce detectable kSZ effect.

Considering the CGM may account for a significant fraction of the missing 
baryons of galaxies, we here present a work to probe the prevailing CGM 
through the measurement of the kinetic SZ effect at the galaxy positions in the 
CMB map. Previous work by \citet{Singh2015} has theoretically studied the 
tSZ/kSZ effect of the CGM, and discussed the detectability of the SZ effect of 
the CGM as well as the constraints on the gas fraction. We here alternatively 
propose to measure the kSZ effect of the CGM and the detectability of the 
signal with different profiles. We try to design an unbiased minimum variance 
estimator of the kSZ effect of the CGM in order to minimize the noise, and then 
stack the velocity weighted kSZ signal of galaxies to further improve the 
measurement. The minimum variance estimator involves finding out an optimal
filter to filter out the large scale contamination of the primary CMB 
while keep the kSZ signal of CGM conserved. The design is unprejudicedly 
constructed for both the ICM in galaxy groups and the CGM around 
galaxies, provided we have the knowledge of the profile of the optical 
depth profile of the ICM/CGM \citep[e.g. see][for the application to mock 
sample of clusters]{Li2014}. {\it In particular, we here focus on the kSZ 
effect of the CGM of Milky Way(MW)-size halos.} Since the baryonic budget 
and the radial extent of the CGM are still under debate \citep{Anderson2011, 
Gupta2012, Fang2013}, we here confine the CGM to two compelling profiles: the 
hot extended halo profile \citep[][hereafter MB profile]{MB2004} and the
empirical $\beta$ profile \citep{Cavaliere1976}. In this work, we are mostly
interested in the relatively flatter MB profile, as extensively discussed in
\citep*{Fang2013} concerning the Milky Way, and it proves a promising profile
of the missing baryons of the MW-size halos. We treat the $\beta$
profile as a comparison model, which can give some insight to the upper limit
of detectability of the CGM.

The fiducial cosmology we adopt in this work is the standard flat $\Lambda$CDM 
cosmology \citep{Planck2015XIII}: $\Omega_m=0.308, h=0.678, n_s=0.968,
\sigma_8=0.815, \Omega_b=0.048$, the cosmic baryon fraction 
$f_b=\Omega_b/\Omega_m=16\%$. We focus on the detectability of SKA-like galaxy
surveys and ACT-like SZ surveys, and discuss the possible application to other
surveys in \S \ref{sec:discussion}. SKA \footnote{http://www.skatelescope.org/}
will survey $10^8$ HI galaxies in a wide sky area and a deep volume, and will
make a fair sample of the reservoir of the CGM of late type galaxies. For MW-
size galaxies to be resolved, we need SZ surveys with the resolution like the
ACT/SPT survey.

The organization of the paper is as follows. We first introduce the models and  
show the kSZ effect of the CGM in \S \ref{sec:modeling_CGM_kSZ}, and then raise 
the formalism and methodology to measure the kSZ effect from the CGM in section 
\S \ref{sec:methodology}. We then detail the cross correlating kSZ effect with 
recovered velocity in \S \ref{sec:isolating_kSZ} and further apply the matched 
filter to improve the detection and predict the detectability in \S 
\ref{sec:results}. We present the conclusion and discussion in \S 
\ref{sec:discussion}.

\section{The kSZ effect of the CGM}
\label{sec:modeling_CGM_kSZ}
When CMB photons travel along the line of sight, the high energy electrons in 
the intervening medium (e.g. ICM, IGM or CGM) will scatter off the CMB photons 
and distort the energy spectrum of the CMB via the inverse Compton scattering. 
This process gives rise to the tSZ effect due to the thermal motion of 
electrons, and to the kSZ effect if the scattering medium has a bulk motion 
with respect to the rest frame of the CMB. For the CGM of a Milky Way-size
galaxy, the kSZ effect is larger than the tSZ effect \citep{Birkinshaw1999, 
Singh2015}. Suppose we have a galaxy with its centre at the position 
${\boldsymbol\theta}_0$ in the 2D sky, where we have adopted the flat sky 
approximation as the galaxy's scale is very small 
\citep{Zaldarriaga1997, Bernardeau2011}. The peculiar momenta of the electrons 
in the surrounding CGM will induce the kinetic SZ effect 
$\Theta({\boldsymbol\theta}- {\boldsymbol\theta}_0)$. In this work, we assume 
the kSZ signal of the CGM is spherical symmetry around the galaxy centre, and 
we can always put the galaxy centre ${\boldsymbol \theta}_0$ at the origin,  
therefore from herein we write any function of $\boldsymbol\theta$ as a 
function of the radial separation $\theta$. The induced kSZ signal is 
proportional to the electron momentum integrated along the line of sight
\begin{equation}\label{eq:ksz_definition}
\begin{split}
	\Theta (\theta)\equiv \frac{\Delta T_{\rm kSZ}}{T_{\rm
	CMB}} & =-\int d\chi a \sigma_T n_e(\theta,\chi) \frac{u_\parallel}{c}\\
&=-\int d\chi \frac{d\tau(\theta)}{d\chi} \frac{u_\parallel}{c} .
\end{split}
\end{equation}
$\tau (\theta)= \int d\chi a \sigma_T n_e(\theta,\chi) $ is the optical depth 
profile of the CGM, where $a$ is the scale factor, $\chi$ is the comoving 
radial distance, $n_e$ is the electron number density profile, and $\sigma_T$ 
is the Thompson cross section. 

Since for a single galaxy, the peculiar velocity along the line of sight
$u_\parallel$ is nearly constant, the kSZ induced temperature distortion is
proportional to the product of the optical depth and the line of sight velocity
\begin{equation}
	\Delta T_{\rm kSZ}(\theta)=\tau(\theta) T_{\rm CMB}
	\frac{u_\parallel}{c}= \tilde{u} \tau(\theta).
\end{equation}
Here, we have defined  $T_{\rm CMB}u_\parallel/c$ as $\tilde{u}$. The observed
optical depth profile is the convolution $\tau_b(\theta)\equiv 
(\tau\star B)(\theta)$ between the underlying profile $\tau(\theta)$ and the 
beam window $B$. We assume the beam of the SZ survey is a Gaussian profile 
$B(\theta)=\!\exp{ (\!-\theta^2/2\theta_b^2) } /2\pi\theta_b^2$, with
$\theta_b\equiv 1/l_b=\!\theta_{\rm FWHM}/ \!\sqrt{8\ln{2}}$, and
$\theta_{\rm FWHM}$ the full width half maximum (FWHM) of the beam profile. 
Accordingly, the beam window in Fourier space can be written as 
$B(l)=\exp{(-l^2/2 l_b^2)}$. In practice, we calculate the smoothed optical 
depth profile in Fourier space as $\tau_b(l)=\tau(l) B(l)$.

  \subsection{Modelling the density profile of the CGM}
  \label{subsec:density_CGM}
    The mass fraction and the underlying spatial distribution of the CGM are 
still under debate. Self-regulated galaxy simulations \citep{Guedes2011, 
Moster2011, FG2011} indicate that the baryonic fraction of galaxies can vary 
substantially. A reasonable range covers 30-100\% of the cosmic mean $f_b$. On 
the other hand, observations have made little advance in constraining 
the mass fraction of the CGM, which covers an even broader range of the mass 
fraction $\sim$ 10-100\% \citep{Anderson2011, Dai2012, Humphrey2011, 
Humphrey2012, Werk2014}. The profile of the CGM is also an open question, and 
the empirical $\beta$ profile is often adopted in the analysis of the 
observational data, especially for elliptical and massive spiral 
galaxies \citep{Anderson2011, Anderson2013}. However, \cite{Fang2013} found 
that the MB profile \citet{MB2004} can account for the missing baryons of the 
Milky Way, and is favoured by current observational constraints. In this work, 
we take the MB profile as a fiducial profile of the CGM for MW-size spiral 
galaxies, and the $\beta$ model as a comparison profile. 

    We calculate the kSZ effect in the context of future galaxy surveys such as 
SKA. Such survey will detected a huge number of HI field galaxies and is less 
subject to the contamination of the hot ICM. We expect that most galaxies will 
be located in halos with mass range $10^{12}-10^{13}$\Msun, and assume 
that gas within the virial radius is in the hydrostatic equilibrium.
Given the redshift distribution and mass distribution of the surveyed galaxies, 
we can in principle calculate the differential contribution of host halos in 
certain  mass and redshift bin, and obtain the total stacked signal. Though, we 
will take the example of the typical Milky Way-size halo of $10^{12}$ M$_\odot$ in 
this work.

The radial density profile of the MB model ( also known as extended hot 
halo model) follows
\begin{equation}\label{eq:gas_density}
    \rho_g(x)=\rho_{\rm v} f_{\rm MB}(x)\ ,
\end{equation}
with the profile 
\begin{equation}
 f_{\rm MB}(x)= \left[ 1+\frac{3.7}{x}\ln (1+x) -\frac{3.7}{C_V}\ln (1+C_V)
\right]^{3/2}\ ,
\end{equation}
where $\rho_\mathrm{v}$ is the normalized value of gas density at virial radius 
(Please find details in \citet*{Fang2013}). $x=r/r_s$, and $r_s$ is the
characteristic radius defined by $r_s=r_\mathrm{v}/C_\mathrm{v}$, where the
virial radius $r_\mathrm{v}=260$ kpc and the concentration $C_\mathrm{v}$=12
have been adopted for Milky Way-size host halo. Assuming a baryonically closed 
system, the baryonic mass of the CGM within the virial radius is $ M_{\rm 
gas}\equiv f_{\rm gas} f_b M_\mathrm{v} =\rho_\mathrm{v} r_s^3 
\int_0^{C_\mathrm{v}}\! \mathrm{d}^3 x f_{\rm MB}(x)$.  In the following, we 
assume the baryon fraction $f_b=\Omega_b/\Omega_0$, and the gas fraction 
$f_{\rm gas}=M_{\rm gas}/f_b M_\mathrm{v}$ within the virial radius. We assume 
the missing mass of hot gas is $10^{11} {\rm M}_\odot$ as in their work, which 
implies $f_{\rm gas}=0.625\%$ of the cosmic mean.

For comparison, besides the fiducial profile, we also test the detectability of 
a steeper density profile, the $\beta$ profile, which follows
\begin{equation}
 f_{\beta}(x)=\left(1+ x^2 \right)^{-3\beta/2}  .
\end{equation}
The $\beta$ profile usually fits well to the hot gas around elliptical galaxies 
\citep{Forman1985} and galaxy groups and clusters\citep{Sarazin1986}.
A realistic $\beta$ profile would require fitting two parameters, the core 
radius $r_c$ and the profile index $\beta$, to the X-ray emission data. We here 
assume the index $\beta=2/3$, and fix $r_c=r_s$ for a direct comparison between 
the MB profile and the $\beta$ profile. We also require a density 
normalization that integrated gas mass within the virial radius to be $f_{\rm
gas} f_b M_{\rm v}$. Though the $\beta$ profile is often adopted for the hot gas
around elliptical galaxies and ICM in galaxy clusters, it has been applied to
giant spiral galaxies lately \citep{Anderson2011, Dai2012}. Whether $\beta$ 
profile is appropriate for CGM in late type galaxies is still under debate 
\citep{Anderson2013}. We here choose the $\beta$ profile to demonstrate the
detectability of the CGM, which may put constraints on its upper limit.

\subsection{The kSZ effect of the CGM}
\label{subsec:ksz_of_CGM}
The profile of the optical depth can be obtained by integrating the gas density 
alone the sightline : 
\begin{equation}
\begin{split}
	\tau(\theta) &\equiv \tau(r_p/\chi(z))= 2 \int_0^{r_m}\mathrm{d} l 
\sigma_T n_e(\sqrt{l^2+r_p^2}) \\
	=&2a f_g \chi_i\frac{\sigma_T\rho_\mathrm{v} r_s}{\mu_e m_p} \int_0^{x_m}
\!\mathrm{d}x f(\sqrt{x^2+x_p^2}) \\
	=& 6.5\times 10^{-5}  \left(\frac{f_b f_{\rm gas}}{0.1}\right) \left(
\frac{1}{1+z}\right) 	\left(\frac{0.26 {\rm Mpc} }{r_{\rm v}} \right)^2 \\
	&\vspace{2ex} \left( \frac{M_{\rm v}}{10^{12} {\rm M}_\odot} \right)
\left( \frac{\int_0^{x_m} \mathrm{d}x f(\sqrt{x^2+x_p^2})}{20} \right) ,
\end{split}
\label{eq:optical_depth}
\end{equation}
where $\chi_i$ is the ionization fraction, and $r_p$ is the projection length. 
At low redshift, we assume the universe is fully ionized such that 
$\chi_i=1$. $r_p=\chi \theta$ is the projected distance of the line of sight to 
the centre of the galaxy, and $x_m=\sqrt{(C_\mathrm{v}^2-x_p^2)}$ is the upper 
range of the integration which means we cut the integration at the virial radius. 
$\mu_e=1.14$ is the electron molecular weight, and $m_p$ is the proton mass.
The observed optical depth $\tau_b$ is the convolution of the underlying optical
depth $\tau$ and the beam profile $B$, i.e. $\tau_b(\theta)=(\tau \star B)
(\theta)$.

\begin{figure}
\includegraphics[width=3.2in]{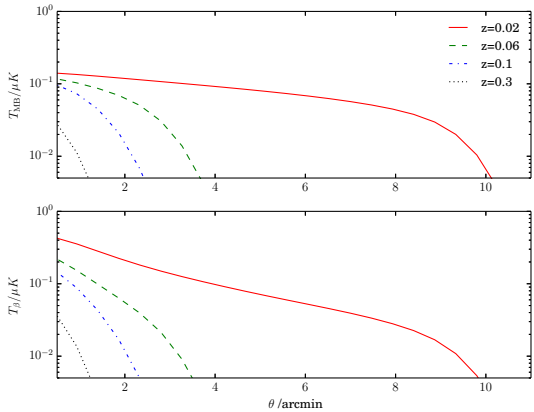}
\caption{The beam convolved kSZ effect profile of a typical MW-size halo at 
redshifts 0.3,0.1,0.06 and 0.02 for both MB profile (upper panel) and $\beta$ 
profile (lower panel). The kSZ signal of MB model is much flatter in the entire
region, and smaller than $\beta$ profile in the central region. 
\label{fig:tksz}}
\end{figure}

For a MW-size halo with a virial mass $M_\mathrm{v} = \times 10^{12}
{\rm M}_\odot$, the typical virial radius is $\sim 0.5$ arcmin at $z\sim 0.3$. It can
be marginally resolved by the ACT-like SZ survey. At higher redshift, the
galaxies are not resolved or just marginally resolved so that most of the 
signal is within the beam and concentrated in the central pixel regardless of 
spatial profile. Thus a similar optical depth profile for different CGM 
profiles is expected after the beam convolution at higher redshifts. Below 
redshift $z\sim 0.3$, the galaxy can be well resolved into several concentric 
rings of pixels. The beam convolved kSZ profiles of the CGM are shown in Fig.
\ref{fig:tksz} at several redshifts ($z=0.02, 0.06,0.1,0.3$). At redshifts 
$z<0.1$, the kSZ profile of the MB model is much flatter and the signal 
amplitude is much smaller than those of the $\beta$ profile in the central 
region. We expect less benefit from filtering the MB's profile, because the 
filter may not well separate the spatial pattern between the kSZ effect of 
extended hot halo and the primary CMB.

\section{Methods and formalism}
\label{sec:methodology}
The kSZ effect is immersed in the overwhelming contamination constituted by the 
thermal SZ effect $\Delta T_{\rm tSZ}$, the primary CMB fluctuation $\Delta 
T_{\rm CMB}$, the foreground cosmic infrared background $\Delta T_{\rm CIB}$ 
\citep{Ade2011, Reichardt2012}, and the instrument noise $T_{\rm det}$ 
imposed by the detector. The noisy temperature distortion profile $\Delta 
T(\theta)$ of the CGM at galaxy position is given by
\begin{equation}\label{eq:noisy_map}
\begin{split}
 \Delta T(\theta) &=\Delta T_{\rm kSZ}(\theta) + \Delta T_{\rm tSZ}(\theta) + 
\Delta T_{\rm CMB}(\theta) + \Delta T_{\rm CIB} (\theta) \\
 	&+T_{\rm det}(\theta) \\
 &=\Delta T_{\rm kSZ}(\theta) + N(\theta) ,
\end{split}
\end{equation}
where we have contracted $\Delta T_{\rm tSZ}+ \Delta T_{\rm CMB} + \Delta 
T_{\rm CIB} +T_{\rm det}$ as a term $N$, as they are all noise to the kSZ 
effect. 

In realistic surveys, the sky map is pixelated on the receiver. For each
pixel in the observed CMB map, both sky and the instrument noise contribute to 
the measured temperature. All the components except the instrument noise 
$T_{\rm det}$ should be smoothed by the telescope beam. To model
the instrument noise, we assume that errors are uncorrelated between pixels and
have uniform variance $\sigma^2_p$, i.e. $\langle T_{\rm det}^i T_{\rm det}^j
\rangle=\sigma^2_p \delta_{ij}$ \citep{Knox1995}. The power spectrum of the
instrument noise thus can be written as
\begin{equation}
 C_l^{\rm det}=\sigma_p^2 \theta_p^2
\end{equation}
For ACT-like survey, we adopt $\theta_{\rm FWHM}=1.4$ arcmin for the beam size, 
pixel size $\theta_p=0.5$ arcmin, and $\sigma_p=2\mu$K noise per pixel. These 
fiducial parameters are adopted through the paper unless specified otherwise.

The problem of probing the kSZ effect of the CGM is now equivalent to the
problem of optimally solving for $\Delta T_{\rm kSZ}$ in Eq.
\ref{eq:noisy_map}.  The basic idea is outlined as follows. The problem
requires separating the kSZ signal from the noisy CMB map by constructing a
correlation, and then beating the noise from all the other components, mainly
from the primary CMB.  As shown in previous works, the kSZ effect of the CGM is
correlated with an effective velocity field $v$, which we can reconstruct from
spectroscopic galaxy surveys. The correlation is only contributed by kSZ
effect, but the noise term is contributed by all the components in the CMB
sky plus the instrument noise. We can then optimally remove the noise with the
matched filter. This would require the knowledge of the optical profile of the
underlying kSZ signal of the CGM. Finally we stack galaxies of the same host
halo mass to boost the signal to noise $(S/N)_{N_{\rm gal}}\propto \sqrt{N_{\rm 
gal}}$.

\subsection{Isolating the kSZ effect of the CGM from the background sky}
\label{subsec:first_velocity}
The presence of the overwhelming noise, e.g. the primary CMB, the CIB from 
dusty star forming galaxies, the tSZ from the very source of kSZ effect, and 
the potential correlation between the CIB and the tSZ, makes the direct 
detection of the kSZ effect nearly impossible. In principle, one can separate 
the signal if there are different spectral dependences between different 
components of the noise term, by taking advantage of modern CMB surveys with 
multiple bands. However, the kSZ effect itself is not only much smaller than 
other components, but is also degenerate with the primary CMB, as it lacks 
spectral dependence.

The local velocity or momentum is a good tracer of the underlying kSZ effect.
We can therefore weight the temperature distortion profile $\Delta T(\theta)$ 
around galaxies with an effective velocity ${\bf v}$ at the galaxy position.
Galaxy density $\delta_g$ traces the underlying dark matter distribution 
$\delta$ within an accuracy of a bias factor, i.e. 
$\delta_g=b_g \delta$. With the aid of spectroscopic galaxy survey, we can 
reconstruct an effective velocity field ${\bf v}$ that traces the underlying 
peculiar velocity ${\bf u}$, by assuming that the matter density $\delta$ 
follows the linear growth theory. Such approximation was widely adopted, and 
the reconstructed velocity traces the underlying velocity well \citep[see][for 
more discussion]{Ho2009, Shao2011b, PlanckIntResCV}. Given the galaxy 
distribution, we can construct the underlying dark matter density field 
$\delta$, which should follow the continuity equation of dark matter in the 
linear approximation
\begin{equation}\label{eq:continuity_equation}
  \dot{\delta} +\nabla \cdot (1+\delta) {\bf u} =0  \ ,
\end{equation}
where ${\bf u}$ is the underlying peculiar velocity. In the regime of the 
linear perturbation theory, one can reconstruct the effective velocity field 
${\bf v}$ with 
\begin{equation}
	{\bf v}({\bf k})=\frac{i a H(z) f \delta_g(k)}{b_g k^2} {\hat{\bf k}} ,
\label{eq:vk_reconstruction}
\end{equation}
where $H(z)$ is the Hubble parameter at redshift $z$, $f=d\ln{D}/d\ln{a}$ and 
$D$ is the linear density growth rate. We have employed that the density follows
the linear perturbation theory, and the galaxy is linearly biased against the 
dark matter with $\delta_g=b_g \delta$, where $b_g$ is the linear bias of 
galaxies. From galaxy surveys, we can measure the galaxy bias $b_g$ beforehand. 
Though there are stochasticities in galaxy bias, studies have shown that the 
uncertainty in galaxy bias is likely to be at the level of 10\% 
\citep{Bonoli2009, Baldauf2010}, and can be negligible at the scales of 
interest \citep{Shao2011b}.

This construction cancels out the cross correlation between the velocity and 
all the other components except the kSZ effect, since they don't have the 
characteristic directional dependences and should be uncorrelated with the 
reconstructed velocity. Details can be found in \citet{Shao2011b}. Therefore, 
we are able to isolate the kSZ signal of the CGM. Statistically, the line of 
sight component of the recovered velocity $v_\parallel$ is correlated with 
underlying velocity field $u_\parallel$, and is thus correlated with the kSZ 
effect $u_\parallel \tau_b$ (momentums).  So that we obtain the correlation of 
the recovered velocity and the CMB map 
\begin{equation}\label{eq:stacking_corr}
\begin{split}
  \langle v_{\parallel} \Delta T(\theta)\rangle &
    =\left< v_{\parallel} \left[ \Delta T_{\rm kSZ}(\theta) + N(\theta) \right] 
\right> \\
    &=\langle v_{\parallel} T_{\rm kSZ}(\theta) \rangle \\
    &=\langle v_{\parallel} \tilde{u} \rangle  \tau_b (\theta) .
\end{split} 
\end{equation}
In the second line of the equation, we have dropped out $\langle v_{\parallel} 
(\Delta T_{\rm tSZ}+\Delta T_{\rm CMB} +\Delta T_{\rm CIB} + \Delta T_{\rm 
det}) \rangle$. When we stack the correlation, we assume that all the galaxies 
share the same optical depth profile $\tau_b(\theta)$ (see the discussion below
for the possible uncertainty of $\tau$ profile).

Now that one separates the kSZ signal from the noisy map by constructing the 
correlation estimator between the reconstructed velocity and the underlying kSZ 
signal of the CGM, one need to assess the performance of the estimator. One of 
the most important statistics to evaluate the performance is the measurement 
error, which is known as the variance of the correlation $\langle v_\parallel 
\Delta T(\theta) \rangle$,
\begin{equation}\label{eq:stacking_var_nofilter}
\begin{split}
 {\rm Var}&=\langle [v_{\parallel} \Delta T(\theta)-\langle v_{\parallel} 
\Delta T(\theta)\rangle ]^2 \rangle \\
    &=\langle [v_\parallel \Delta T(\theta)]^2 \rangle- \langle v_\parallel 
\Delta T(\theta) \rangle^2	\\
    &\simeq \langle v^2_\parallel \rangle \langle \Delta T^2(\theta) \rangle 
\ .
\end{split}
\end{equation}
In the last line of Eq. \ref{eq:stacking_var_nofilter}, we omit $\langle 
v_\parallel \Delta T(\theta) \rangle^2$ since it's much smaller than the 
variance $\langle (v_\parallel \Delta T(\theta))^2 \rangle$. We also made an 
assumption that $v_\parallel$ and $\Delta T(\theta)$ follow the distribution of 
Gaussian fields. We then expand the equation using Wick theorem, and omit any 
terms concerning 3-point correlation.

The measurement error is constituted by the product of the velocity 
dispersion $\langle v_\parallel \rangle$ and the temperature variance $\langle 
\Delta T(\theta) \Delta T(\theta)\rangle$ of each pixel in the CMB survey.
It's clear from Eq. \ref{eq:stacking_var_nofilter} that the temperature 
variance of each pixel is contributed by all the components in the pixel, 
including the predominant CMB variance. Therefore, cross correlating the 
CMB map with the reconstructed velocity $v_\parallel$ merely separates the kSZ
effect of CGM from other components of the noisy CMB map, while the detection
is still subject to large measurement error.  As shown later in \S 
\ref{subsec:signal_to_noise_nofilter}, stacking $\sim$ 1 million MW-size halos 
is necessary to achieve 1 $\sigma$ detection.

Thus a direct cross correlation between the kSZ effect of the CGM with the 
reconstructed velocity picks out only the kSZ effect. To improve the detection, 
we need to further reduce the measurement error by removing the contamination 
of all the other components. This involves utilizing the knowledge of the 
covariance matrix of the noise, and the spatial distribution of the kSZ signal 
of the CGM, which in turn lies in the optical depth profile $\tau_b(\theta)$ as in 
Eq. \ref{eq:stacking_corr}. Given the profile of $\tau$, we can design an 
unbiased estimator $U$ of the underlying velocity field $\tilde{u}$ with a
matched filter $\phi$ to optimally remove the noise mainly from the primary
CMB. Then by cross correlating $\langle v U\rangle$ one can achieve a much 
higher S/N.

\subsection{Matched filter for the underlying kSZ signal of the CGM}
We now turn to the problem of finding out an optimal estimator of $\Delta
T_{\rm kSZ}$ from Eq. \ref{eq:noisy_map}. The primary CMB dominates the 
variance in each pixel. As shown in \S \ref{subsec:signal_to_noise_nofilter},
the primary CMB gets the most of the contribution from large scale multipoles, 
while the kSZ signal from the CGM is mainly from the scale of galaxy size. Thus 
we can take advantage of the different scale dependences and optimally subtract 
the primary CMB from large scales with an unbiased minimum variance estimator 
$U$, by employing matched filter \citep[e.g.][]{Haehnelt1996, Tegmark1998, 
Mak2011, Li2014}. An optimal matched filter $\phi(\theta)$ should satisfy the 
following requirements: (1) $U$ is an unbiased estimator of the amplitude 
$\tilde{u}$ of the underlying kSZ signal; (2) The variance between the 
estimator $U$ and underlying signal $\tilde{u}$ is minimized.

The nominal temperature distortion profile around one galaxy is $\Delta 
T(\theta)= \Delta T_{\rm kSZ}+ N(\theta)= \tilde{u} \tau_b(\theta)+N(\theta)$. 
We can construct an effective quantity $U$ as an estimate of $\tilde{u}$,
\begin{equation}
 U=\int d^2\theta \phi(\theta) \tilde{u} \tau_b(\theta)+\int d^2\theta
\phi(\theta) 
N(\theta)\ ,
\end{equation}
where $\theta$ the radial separation from the galaxy centre and $\phi(\theta)$ 
is the matched filter. The first condition requires an unbiased estimator, 
thus the expectation of $U$ should be $\tilde{u}$
\begin{eqnarray}
 \langle U\rangle= \langle \tilde{u}\rangle \int\!\mathrm{d}^2\theta 
\phi(\theta)
\tau_b(\theta) +\int\!\mathrm{d}^2 \theta \phi(\theta)\langle  N(\theta) 
\rangle ,
\end{eqnarray}
which means
\begin{equation}
 \int\!\mathrm{d}^2 \theta \phi(\theta) \tau_b(\theta)=1 \ .
\end{equation}
In the above equation, we have used the property that the ensemble average of 
the noise term $N$ vanishes.

The second condition of the estimator requires minimizing of the variance
$\langle (U-\tilde{u})^2 \rangle$ between the estimator and the underlying
amplitude. We can write the variance in Fourier space using Parseval theorem
\begin{equation}
\begin{split}
 \langle (U-\tilde{u})^2 \rangle&=\langle \left[\tilde{u} \int d^2 \theta 
\phi(\theta) \tau_b(\theta)
  +\int\!\mathrm{d}\theta \phi(\theta)) N(\theta)\right]^2 \rangle -
\langle \tilde{u} \rangle ^2 \\
  &= \frac{1}{(2\pi)^4} \langle \left[ \int\!\mathrm{d}^2 l \phi(l) 
N(l)^{\ast} \right] \left[\int \!\mathrm{d}^2 l^{\prime} 
\phi^{\ast}(l^{\prime}) N(l^{\prime} )\right] \rangle \\
  &= \frac{1}{(2\pi)^2} \int\!\mathrm{d}^2 l \int\!\mathrm{d}^2 l^{\prime} 
\delta^D(l-l^{\prime})
(C_l B_l^2+\sigma_p^2 \theta_p^2) \phi(l) \phi^{\ast} (l^{\prime}) \\
  &= \frac{1}{(2\pi)^2}\int\!\mathrm{d}^2 l (C_l B_l^2 +\sigma_p^2 \theta_p^2)\,
\left|\phi(l)\right|^2 \ ,
\end{split}
\end{equation}
in the following transform convention
\begin{equation}\label{eq:FT_convention}
\begin{cases}
	\phi(l) &= 2\pi\! \displaystyle\int \phi(\theta) J_0(l\theta) \theta 
\: d\theta 	
\\
	\phi(\theta) &= \frac{1}{2\pi}\! \displaystyle\int \phi(l) J_0(l 
\theta) l\: dl ,
\end{cases}
\end{equation}
which has been reduced to the zero-order Hankel transform due to circular 
symmetry. $C_l$ is the angular power spectrum of the primary CMB, for which we 
have used the CAMB\footnote{http://camb.info} code \citep{Lewis2000CAMB, 
Lewis2002}. We have assumed that there's no correlation between the primary CMB 
and the instrument noise.

Therefore, solving the matched filter can be reduced to finding the local 
extremum of a function that is subject to an equality constraint. Specifically, 
here we need to solve for $\phi$ and a Lagrangian multiplier $\lambda$ that 
minimize
\begin{equation}
 \left[ \int\!\mathrm{d}^2 l (C_l B_l^2 + \sigma_p^2 \theta_p^2)
|\phi|^2(l)-\lambda 
(\int\!\mathrm{d}^2 l \phi^{\ast}(l) \tau_b(l)-1) \right] ,
\end{equation}
with the constraint 
\begin{equation}
\label{eq:unbiased_Fourier}
 \int\!\mathrm{d}^2 \theta \phi(\theta) \tau_b(\theta)= \frac{1}{(2\pi)^2} 
\int\!\mathrm{d}^2 l \phi(l)\tau_b^{\ast}(l)=1 .
\end{equation}
Hence the following equation is demanded
\begin{equation}
\label{eq:lambda_Fourier}
 \int\!\mathrm{d}^2 l \delta\phi^{\ast}(l) \left[ (C_l B_l^2+\sigma_p^2
\theta_p^2) \phi(l) -\lambda \tau_b(l) \right] =0 \ .
\end{equation}
We thus have the solution by solving Eq. \ref{eq:unbiased_Fourier} and 
\ref{eq:lambda_Fourier}
\begin{equation}
\label{eq:phi_Fourier}
 \phi(l)=\lambda \frac{\tau_b(l)}{C_l B_l^2+\sigma_p^2 \theta_p^2}\ ,
\end{equation}
where the normalization $\lambda$ is given by
\begin{equation}
 \lambda=\left[ \int\!\mathrm{d}^2 l \frac{\tau_b(l)\tau_b^{\ast}(l)}{C_l 
B_l^2+\sigma_p^2 \theta_p^2}\right]^{-1} \ .
\end{equation}

We then apply this filter to the noisy CMB map, and the cross correlation 
signal would be 
\begin{equation}
\begin{split}
 \langle v_\parallel U\rangle&=\langle v_\parallel \int d^2 \theta \phi(\theta)
\left[\Delta T_{\rm kSZ} +N(\theta) \right]^2 \rangle \\
  &=\langle v_\parallel \tilde{u} \rangle \left( \int d^2\theta
\phi(\theta) \tau_b(\theta)\right) \\
  &=\langle v_\parallel u_\parallel \rangle T_{\rm CMB}/c .
\end{split}
\end{equation}
By constructing the correlation of the optimal estimator and the recovered
velocity, we expect not only to get the kSZ effect isolated from the
noisy map, but also to have a greatly reduced measurement error. By stacking
galaxies with the same host halo mass, we will further beat the statistical
noise and get a higher S/N.

\section{Isolating the kSZ signal of the CGM from the CMB sky}
\label{sec:isolating_kSZ}

  \subsection{The signal to noise ratio without filtering}
  \label{subsec:signal_to_noise_nofilter}
If no filter is applied to the noisy CMB map, emission from all scales of every 
component in the CMB sky will be retained in the measurement error. From Eq.  
\ref{eq:stacking_corr} and \ref{eq:stacking_var_nofilter}, we can now estimate 
the signal to noise ratio of this correlation,
\begin{equation}\label{eq:signal_to_noise_nofilter}
 \begin{split}
  \left(\frac{S}{N}\right)^2&=\frac{\left< v_\parallel u_\parallel\right>^2
\left( T_{\rm CMB} \tau_b(\theta)\right)^2} {\langle v_{\parallel}^2 \rangle 
\langle \Delta T^2(\theta)\rangle}  \\
   &= \frac{ r_{uv}^2 \langle u^2_\parallel \rangle
   \left[ T_{\rm CMB} \tau_b(\theta)\right]^2 }{\langle \Delta
T^2(\theta)\rangle} \\
  &= r_{uv}^2 \frac{\langle \Delta T^2_{\rm kSZ}(\theta) \rangle}{\langle 
\Delta T^2 \rangle}
 ,
 \end{split}
\end{equation}
where $r_{uv}$ the correlation coefficient defined as $r_{uv} \equiv \langle 
v_\parallel u_\parallel \rangle / \sqrt{\langle v^2_\parallel \rangle \langle 
u^2_\parallel \rangle}$. Thus the signal to noise ratio
of each pixel is essentially proportional to the correlation coefficient $r_{uv}$
and the ratio of the variance of the kSZ effect to the overall variance of the
CMB sky.

All the components in the CMB map constitute the pixel noise to the kSZ 
signal, and it can be obtained from the following equation
\begin{equation}
\label{eq:pixel_variance}
  \langle \Delta T(\theta) \Delta T(\theta) \rangle=\frac{1}{(2\pi)^2}\int 
\mathrm{d}^2 l C_l^{\rm tot} B^2(l)\ .
\end{equation}
$C_l$ is the measured power spectrum of the CMB sky, which is the sum of the
primary CMB, the SZ effect, the CIB, etc. . However, the primary CMB dominates 
on scales $l<1000$ and gets most of its variance around the first several
acoustic peaks. The power spectra of other components will only exceed CMB on
smaller scales, and get suppressed by the beam window of the telescope, and
thus contribute negligible power to the pixel variance. Hence we calculate the
pixel variance in Eq. \ref{eq:pixel_variance} with only the primary CMB, which
we get from the CAMB code. The resulting pixel variance of the temperature
fluctuation is $\sim 12000 \mu$ K$^2$, which means $\sim 110 \mu$ K
of the r.m.s. noise in a pixel.

  \subsection{velocity correlation}
  \label{subsec:velocity_dispersion}
  In the context of the standard cosmology and the linear perturbation theory, 
both baryons and dark halos emmersed in the same gravitational potential 
exerted by the large scale density perturbation and shall follow the same  
velocity field on large scales. In linear regime and even quasilinear regime, 
one can reconstruct an effective velocity field using the continuity equation, 
given a galaxy density field from observations. On small scales, shell crossing 
will complicate the dynamics and fail the linear prediction. Nevertheless, the 
velocity field reconstructed from spectroscopic galaxy surveys can trace the 
underlying velocity flow on large scales. Hydrodynamic plus N-body simulations 
have examined the relation between the underlying velocity and the 
reconstructed velocity. It has been shown \citep{Ho2009, Shao2011b, 
PlanckIntResCV} that the recovered velocity field ${\bf v}$ follows the 
underlying baryonic velocity ${\bf u}$ fairly good. For simplification we here 
denote $u$ and $v$ as the projected velocity component 
along any given direction of the 3D velocity ${\bf u}$ and ${\bf v}$.  To 
quantify the performance of the velocity reconstruction, one usually 
introduces in the Fourier space the cross correlation coefficient $r(k)= 
\Delta^2_{uv}(k)/ \sqrt{\Delta^2_v(k) \Delta^2_{u}(k)}$, and the 
velocity bias $b_{v}(k)=\sqrt{\Delta^2_v(k)/ \Delta^2_u(k)}$ \citep[see 
e.g.][for a different but related definition of the coefficient 
$r(k)$]{Ho2009}, where the power spectrum $\Delta^2_{uv}=P_{uv}(k)k^3/2\pi^2$, 
and $P_{uv}(k)$ is defined as $(2\pi)^3 \delta(k-k^{\prime}) P_{uv}(k)=\langle 
v({\bf k}) u({\bf k}^\prime) \rangle$.

\begin{figure}
 \includegraphics[width=3.2in]{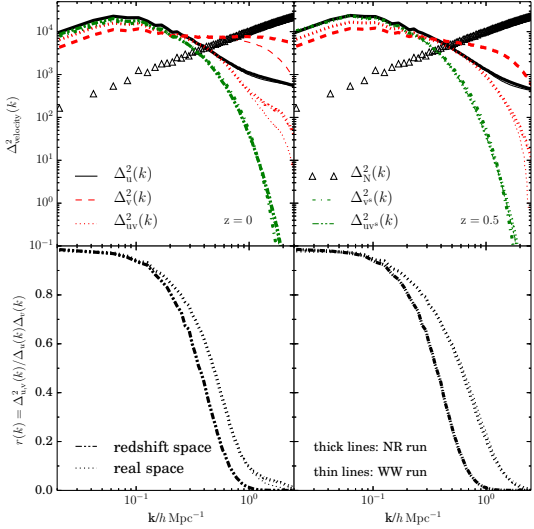}
 \caption{{\it The upper two panels}: The power spectra of velocity at z=0 
(top left) and z=0.5 (top right): $\Delta^2_{u}$ for the underlying baryon 
velocity, $\Delta^2_{v}$ for the recovered velocity from real space matter 
density, $\Delta^2_{uv}$ for the cross power spectrum between the two. The 
recovered velocity power spectrum and the cross power spectrum, given matter
density in redshift space, are shown as $\Delta^2_{v^s}$ and $\Delta^2_{u
v^s}$. The shot noise power spectrum of velocity, which has been scaled to the
fiducial galaxy number density, is shown in triangles. {\it The lower two
panels}: the cross correlation coefficient $r(k)$ between the underlying baryon
velocity $u$ and the recovered velocity given galaxy density in real space and
in redshift space.}
 \label{fig:cor_uv}
\end{figure}

\begin{figure}
  \includegraphics[width=3.2in]{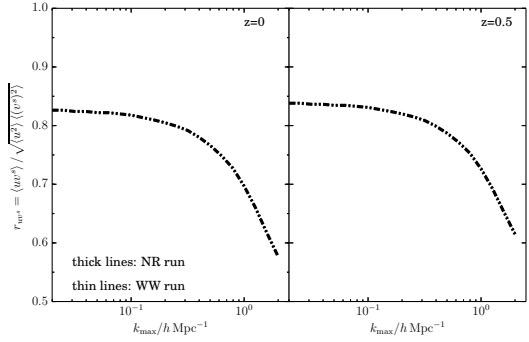}
 \caption{The cross correlation coefficient in the redshift space $r_{uv}$ as a 
function of $k_{\rm max}$ at z=0 (left panel) and z=0.5 (right panel), defined 
as $r_{uv}=\langle u_\parallel v^s_\parallel \rangle/\sqrt{\langle 
u_{\parallel}^2 \rangle \langle (v^s_{\parallel})^2 \rangle}$. The cross 
correlation  and auto correlation are integrated up to $k_{\rm max}$ in the 
convention of Eq. \ref{eq:cross_correlation}. Thick and thin lines denote 
the NR run and WW run respectively. Here $r_{uv}$ encodes the uncertainties 
induced by the redshift space distortion, the gas physics, and the shot noise. 
} 
\label{fig:coef}
\end{figure}

A perfect reconstruction of the underlying velocity means $r(k)\simeq 1$ 
and $b_{v}(k)\sim 1$ at all scales. In realistic galaxy surveys, there are a
number of issues that will complicate the velocity reconstruction 
\citep{Shao2011b}, so that $r(k)$ and $b_{v}(k)$ both deviate from unity. 
However, deterministic bias does not affect the estimation of S/N as
clearly shown in Eq. \ref{eq:signal_to_noise_nofilter}, leaving the dependence
of S/N only on $r_{uv}$. Hence we focus on $r_{uv}$ to evaluate the
reconstruction. Several important issues which may degrade the velocity 
reconstruction are listed here: (1) Eq. \ref{eq:vk_reconstruction} only holds 
when $\delta_g \ll 1$, and we have also simplified the equation by assuming the 
linear density evolution, and a deterministic bias between the galaxy density 
$\delta_g$ and the matter density $\delta$; (2) the galaxy number density in the 
redshift space is what is measured in galaxy survey, while the 
equation holds in the real space; (3) in the reconstruction we lose the 
information of velocity vorticity which contributes significantly to the kSZ 
signal on small scales; (4) the lack of knowledge of the underlying gas 
dynamics will induce uncertainties in the recipes of gas dynamics employed in 
simulation and thus in the reconstructed velocity field; and finally (5) the 
mask and geometry of galaxy surveys will reduce the effectiveness of the 
reconstruction. These factors will of course degrade the reconstruction. 
Fortunately, the velocity correlation length is $\sim 20-40$ \Mpch, and the 
reconstruction is fairly good on linear scale and even on quasilinear scales. 

We test several issues mentioned above against a set of hydrodynamic
simulations. The simulations are run with the GADGET2 code \citep{Springel2005}
in a LCDM cosmology with parameters: $\Omega_m=0.279, \Omega_\Lambda=0.721,
\Omega_b=0.0463, h=0.7$.  The box size of the simulation is L = 300 \Mpch on
each side, in which $768^3$ dark matter particles and $768^3$ gas particles are
initially seeded.  Two runs are implemented with the same initial condition,
one of which is non-radiative (NR run), and the other radiative (WW run) with
star formation and weak wind feedback (wind speed is 480 km/s). We measure the
power spectra of the underlying velocity of baryons $\Delta^2_u(k)$, the
recovered velocity $\Delta^2_v(k)$ given dark matter distribution, the cross
power spectrum $\Delta^2_{uv}(k)$, and the corresponding power spectra in the
redshift space, and show them in the upper panels of Fig. \ref{fig:cor_uv}.
It's shown that the recovered power spectrum of the velocity in the redshift space
is larger than that in the real space on large scales, and on scales $k>0.2$
\hMpc the power spectrum in redshift space drops rapidly due to the Finger of
God effect. Also shown is the shot noise power spectrum, which has been scaled
to the fiducial galaxy number density $1\times 10^{-3}$\hMpct, and it will
dominate the velocity power spectrum at $k>0.3$\hMpc. Comparing the NR run
(thin lines) and WW run (thick lines), we find gas physics only modify small
scale power spectrum. We also show the correlation coefficient in Fourier space 
$r(k)$ in the lower two panels of Fig. \ref{fig:cor_uv}. The cross correlation
coefficient is large than 0.9 on scale $k<0.1$ \hMpc, and decreases rapidly on
smaller scales when nonlinear evolution dominates, and redshift space
distortion further reduces the coefficient. Comparing NR run and WW run, we
find gas physics only affects the reconstruction on scales $k>1$ \hMpc.

  \subsection{Uncertainties in $r_{uv}$}
  \label{subsec:uncertainties_in_r}
  All of above factors play a role in shaping the recovered velocity. Though
there are indeed significant deviations between the power spectra of recovered
velocity and the underlying velocity, what we really measure is the cross
correlation of the recovered velocity and the CMB distortion at galaxy
positions. Thus to quantity the performance of the reconstruction, we measure
the cross correlation coefficient $r_{uv}=\langle u_\parallel v_\parallel
\rangle/ \sqrt{\langle u_\parallel^2 \rangle \langle v_\parallel^2 \rangle} $. The cross
correlation between the reconstructed velocity and the underlying velocity along
the line of sight (L.O.S) can be calculated by integrating the cross power
spectrum as 
\begin{equation}
\begin{split}
  \langle u_\parallel v_\parallel \rangle&=\int_0^{k_{\rm max}}
d \ln k\, \Delta^2_{ u_\parallel v_\parallel}(k) .
\end{split}
\label{eq:cross_correlation}
\end{equation}

The cross correlation converges rapidly to the auto correlation of the
underlying velocity. On one hand, linear theory predicts that peak contribution
comes from $k$ around $0.05$\hMpc (see Fig. \ref{fig:cor_uv}), and 70\% (90\%) 
of the velocity dispersion is contributed on scales $k<0.1$\hMpc ($<0.3$\hMpc). 
On the other hand, observations and simulations have also shown that only 15\% 
of the velocity dispersion comes from $k>0.15$\hMpc, where the linear 
approximation holds very well \citep{Gramann1998, Strauss1995}. We measure 
$\langle u_{\parallel}^2 \rangle$,
$\langle v_{\parallel}^2 \rangle$ and $\langle u_\parallel v_\parallel \rangle$
in the simulation with the convention as shown in Eq.
\ref{eq:cross_correlation}. As shown in Fig. \ref{fig:cor_uv}, most of the
correlation signal comes from large scale modes, which makes sure that the
correlation $\langle u_\parallel v_\parallel \rangle$ converges on large
scales. We then calculate the cross correlation coefficient in the redshift
space $r_{uv}$ as a function of $k_{\rm max}$, and show them in Fig.
\ref{fig:coef}. Here $r_{uv}$ encodes errors induced by the redshift space 
distortion, the gas physics, and the shot noise. We find that the correlation 
coefficient $r_{uv}$ is larger than 0.8 for galaxy surveys sampling a galaxy 
density corresponding to $k_{\rm max}\simeq 0.1$\hMpc, and then decrease to  
below 0.6 at smaller scales when the shot noise becomes overwhelming.  Therefore 
in this work we choose $r_{uv}=0.7$ as the fiducial value, and the signal to 
noise ratio can be scaled accordingly for other values of $r_{uv}$.

We have tested the correlation coefficient $r_{uv}$ against various 
uncertainties in the velocity reconstruction in hydrodynamic simulations,
and we assume $r_{uv}$ applies to our fiducial cosmology. For our fiducial
cosmology, the underlying zero-lag auto-correlation along the line of sight
$\langle u_\parallel^2 \rangle$ is calculated using the linear perturbation 
theory in this work
\begin{equation}
 \langle u_\parallel^2 \rangle=\frac{\left(afH\right)^2}{3} 
 \int_{k_{\rm min}}^{k_{\rm max}} \mathrm{d}\ln k \frac{\Delta^2_m(k)}{k^2} ,
\end{equation}
where $\Delta^2_m(k)$ is the power spectrum of the dark matter obtained from 
CAMB. We have adopted the assumption that the velocity is isotropic. The linear 
prediction of the velocity variance is $\langle u^2_\parallel \rangle \sim 360$ 
km/s in our fiducial cosmology.

As shown in the Eq. \ref{eq:signal_to_noise_nofilter}, the signal to noise
ratio is proportional to the r.m.s of the underlying velocity dispersion and
the correlation coefficient $r_{uv}$. Simple calculation shows that the
central kSZ signal at $z\sim 0.5$ for a MW-size halo is $\lesssim 0.1\mu$K 
(Eq. \ref{fig:tksz}). Therefore, a direct stacking without optimal filtering 
requires more than one million galaxies to achieve 1 $\sigma$ detection. Even with 
more pixels of resolved galaxies at lower redshifts, the required galaxy number 
scales inversely to the number of pixels per galaxy, and is still beyond 
current galaxy surveys for late type galaxies. Compared to the variance of the 
background sky, the instrument noise term is very small. Therefore, the 
measurement error of the CGM's kSZ signal is almost entirely contributed by the 
primary CMB. Thanks to the different scale dependences between the kSZ 
effect and the primary CMB, we can reduce the variance by filtering out the 
contribution of the large scale primary CMB.

\begin{figure}
\includegraphics[width=3.2in]{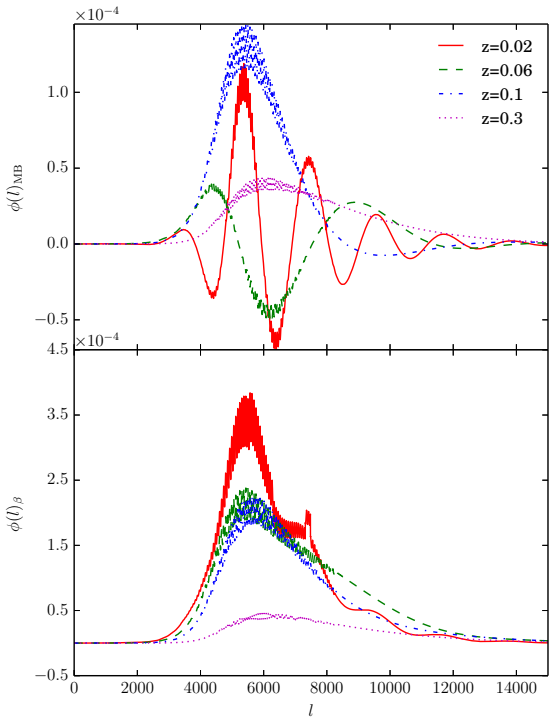}
\caption{The matched filter $\phi(l)$ for extended hot halo model (MB) 
profile and $\beta$ profile in Fourier space. The first peaks of the filters 
are almost at $l\sim$ 5000- 6000. The filters are smooth and they look similar 
for both models at high redshift, while at low redshift the filter for the 
flatter MB model fluctuates much more severely than that for the $\beta$ 
profile.
\label{fig:phil}}
\end{figure}

\begin{figure}
\includegraphics[width=3.2in]{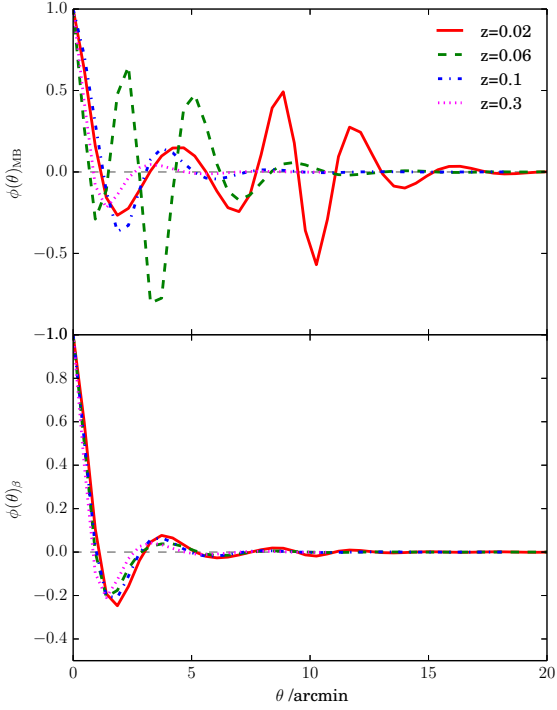}
\caption{$\phi(\theta)$ for ACT-like CMB survey for both MB profile
(upper panel) and empirical $\beta$ density profile (lower panel) for CGM.
\label{fig:phit}}
\end{figure}

\section{The performance of the matched filter}
\label{sec:results}
  \subsection{The profile of the matched filter}
  \label{subsec:matched_filter_profile}
As shown in Eq. \ref{eq:phi_Fourier}, the filter profile is the most 
sensitive to the scales where the signal becomes relatively dominant with
respect to the primary CMB plus the instrument noise. On large scales (low 
$l$), $\phi(l)$ almost vanishes since $\tau(l)$ is much smaller than the 
primary CMB spectrum. However, the primary CMB is damped exponentially, while 
$\tau(l)$ decreases slowly for $l>2000$. In the case of ACT-like CMB survey, 
the filter gets its maximum contribution from those scales where the instrument 
noise is comparable to the primary CMB, i.e. $l\sim 5000-6000$ at almost all 
redshifts. At lower redshifts z$\lesssim$ 0.06, there are several comparable 
troughs and peaks for the MB model, as shown in the Fourier space in Fig. 
\ref{fig:phil}. Because of the acoustic fluctuation in the damping tail of the 
CMB, there's small dithering of the filter. We can also see the behaviour of 
the filter $\phi(\theta)$ in Fig. \ref{fig:phit} in real space, in which we 
have adopted the normalization so that $\phi(\theta=0)=1$. For all the selected 
redshifts shown here, the first trough is located at $\theta \sim 1.5$ arcmin 
away from the galaxy center, which is just around the beam size (the first 
trough of 0-th Bessel function). However, at z$\lesssim$ 0.1, the MB profile is 
very shallow and the kSZ signal from scales larger than the beam size is 
non-negligible, and thus part of the kSZ signal is compromised by the filter. 
That's why one sees that at low redshifts, the filter fluctuates heavily even 
in the outer region of the halo.

We also show in Fig. \ref{fig:phit} the filter for the $\beta$ profile which 
turns out to be similar for all redshifts. As shown in the figure, although the 
spatial extent of the $\beta$ profile (see Fig. \ref{fig:tksz}) increases by an 
order of magnitude from z=0.3 to the local universe, the shape of the filter at
different redshifts changes only slightly. This feature will simplify the
stacking procedure, as the filter follows almost the same shape, and thus apply
to all galaxies at different redshifts.

Also as shown in Fig. \ref{fig:phit}, when galaxies became well resolved by the 
SZ survey, the filter profiles diverge significantly between the two CGM 
profiles. The different behaviours will in turn shed light on how to 
discriminate between the two CGM profiles, as the kSZ effect of a given 
profile only responds to the corresponding matched filter, and gives the 
optimal measurement accordingly.

  \subsection{Evaluating the S/N with filtering}
  \label{subsec:signal_to_noise_filtered}
  The matched filter gives an optimal estimator $U$ of the underlying kSZ 
signal which minimizes the variance. With this estimator in hand, we can cross 
correlate the estimator $U$ and the constructed velocity field $v$, and 
evaluate the signal to noise ratio for a single galaxy as follows
\begin{equation}
\label{eq:signal_to_noise_filtered}
\begin{split}
 &\left(\frac{S}{N}\right)^2  \equiv \frac{\langle v_\parallel 
U\rangle^2}{\langle  (v_\parallel U- 
    \langle     v_\parallel U\rangle)^2 \rangle}  \\
 &=\frac{\langle v_\parallel (\tilde{u} \int\!\mathrm{d}^2
  \theta \,\phi(\theta) \,\tau_b(\theta)+\int\!\mathrm{d}^2\theta\, 
\phi(\theta)\,
 N)\rangle^2} {\langle(v_\parallel U)^2\rangle - \langle v_\parallel u 
\rangle^2} \\
  &=\left[ 
    1+\frac{\langle v_\parallel^2 \rangle \langle \tilde{u}^2 \rangle} 
{\langle v_\parallel \tilde{u} \rangle^2} + 
     \frac{\langle v_\parallel^2\rangle} {\langle v_\parallel 
\tilde{u}\rangle^2}
    \frac{\int\!\int\! \mathrm{d}^2\theta \mathrm{d}^2\theta^{\prime}
  \phi(\theta)\!\phi(\theta^{\prime}) \langle N(\theta)\!N(\theta^{\prime})
\rangle} 
      {\left(\int\!\mathrm{d}^2 \theta\tau_b(\theta)\phi(\theta)\right)^2}
    \right]^{-1} \\
  &= \left[  1+\frac{\langle v_\parallel^2 \rangle \langle \tilde{u}^2 \rangle
}{\langle v_\parallel \tilde{u} \rangle^2}
  + \frac{\langle v_\parallel^2\rangle}{\langle v_\parallel \tilde{u} \rangle^2 
} \frac{ (2\pi)^2\int\mathrm{d}^2 l \phi^2(l)\left[
    C_l B^2(l)\!+\!\sigma^2_p \theta^2_p\right] } {\left[ \int\!\mathrm{d}^2 l
    \phi(l) \tau(l) B(l) \right]^2} \right]^{-1} .  
\end{split}
\end{equation}
We have transformed the estimation into Fourier space in the last line. As
addressed in \S \ref{subsec:uncertainties_in_r}, the signal to noise ratio is
strongly correlated with the cross correlation between the reconstructed
velocity and the underlying velocity. Even if there's a deterministic bias of 
velocity between $v$ and $u$, the bias will cancel out in the denominator and 
the numerator, leaving the dependence only on the underlying velocity 
dispersion and the correlation coefficient $r_{uv}$. We can rewrite the signal 
to noise estimation in the case of matched filtering as
\begin{equation}
\label{eq:signal_to_noise_filtered_rewritten}
\begin{split}
 \left(\frac{S}{N}\right)^2  &= 
	\left[ \!1+\!\frac{1}{r_{uv}^2} \! + \! \frac{c^2}{ r_{uv}^2 \langle 
u_\parallel^2 \rangle} \frac{ (2\pi)^2 \int\mathrm{d}^2 l 
\phi^2(l)\left[ C_l 
B^2(l)\!+\!\sigma^2_p \theta^2_p\right] } {T^2_{\rm CMB} \left[ 
\int\mathrm{d}^2 l \phi(l) \tau(l) B(l) \right]^2 } 
  \right]^{-1} .
\end{split}
\end{equation}
In real survey, even if the contamination is filtered, the residual noise
is still much larger than the kSZ signal of MW-size halos. Thus the
first two terms can be neglected, if the correlation coefficient is good
enough. Therefore, the signal to noise ratio is proportional to the correlation
coefficient $r_{uv}$, and that's why a good velocity tracer is the key quantity
in this method.

\begin{figure}
\includegraphics[width=3.2in]{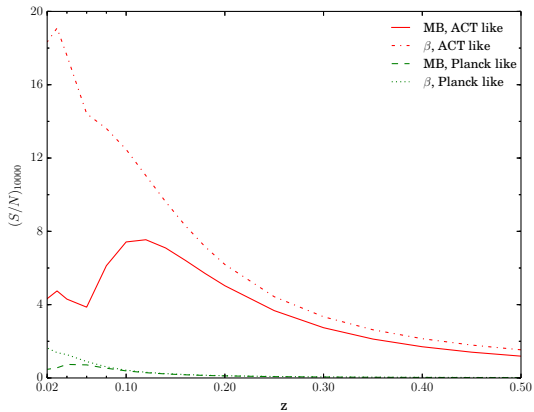}
\caption{The signal to noise ratio $S/N$ for 10000 MW-size halos as a 
function of redshift for an ACT-like SZ survey, for both the MB profile and 
the $\beta$ profile with $\beta=2/3$. Also shown are the predictions for SZ 
surveys like Planck. 
\label{fig:snz}}
\end{figure}

It's clear from Fig. \ref{fig:phil} that $\phi(l)$ will remove the contribution 
from large scale modes of the primary CMB $C_l$. Though the large scale modes of
the kSZ effect $\tau(l)$ are also removed, the loss of signal is negligible and 
the total signal is conserved by design. After applying this filter, the 
contribution of the primary CMB to the noise is greatly reduced, and the r.m.s.
noise is dramatically reduced from $\gtrsim 100\mu$ K to a level of $\lesssim 
5\mu$ K at $z\sim 0.5$. Assuming we stack galaxies of the same host halo mass,
for example 10000 MW-size halos, in a narrow redshift bin around $z\sim0.5$, we
will reduce the statistical noise and achieve $\gtrsim$ 1 $\sigma$ detection. 
At lower redshifts, the CGM will have higher kSZ effect amplitude, and cover 
more extended pixels,  which make the filter much more effective. Therefore, 
the S/N increases until $z\sim 0.15$, and with 10000 MW-size halos one can 
achieve $\sim 7.5 \sigma$ detection.  Below this redshift, the signal to 
noise decreases, as the kSZ effect get non-negligible contribution from 
large scales for spatially extended halos, and thus the filter can not 
effectively suppress the primary CMB on these scales. The signal to noise ratio 
as a function of the redshift for 10000 galaxies in a redshift bin ($\Delta 
z=0.1$ for $z>0.1$, $\Delta z=0.02$ for $z<0.1$) is plotted in Fig. 
\ref{fig:snz}. In fact, we have simplified the estimation here, and the total 
signal to noise ratio will inevitably depend on the number distribution, the 
redshift distribution, and the mass distribution of the host halos. Careful 
calculation is needed when applied to the real galaxy catalogue.

Also shown in the figure is the S/N for the $\beta$ profile, which is much 
steeper in the outer region than the fiducial MB profile. For the $\beta$ 
profile, the mass is more concentrating to the centre, leading to a higher 
kSZ effect in the central region, with little kSZ signal coming from large 
scale wave mode. The kSZ amplitude of the $\beta$ profile is much higher than
the MB profile, and thus we can expect a higher signal to noise ratio at the
same redshift. As shown in the lower panel of Fig. \ref{fig:phil}, the filter
for the $\beta$ profile effectively removes the large scale contribution from
the primary CMB, and retains the kSZ signals very well. Thus the S/N increases 
with decreasing redshift till the turnover redshift, z=0.03 for a $\beta$ 
profile, where we get a maximum S/N=19 and a dramatic drop-off below this 
redshift.

Besides the intrinsic shape of the kSZ profile of the CGM, the beam size of 
the telescope also determine the shape of the filter. For  comparison, we also 
predict the detectability for the two profiles in the case of a Planck-like CMB 
survey, i.e., $\theta_{\rm FWHM}=5$ arcmin, $\sigma_p=5 \mu$K per pixel. The 
results are also shown in the Fig. \ref{fig:snz}. There's similar trends except 
much lower signal to noise ratio. The signal to noise ratio reaches the maximum 
at z$\simeq$ 1 (z$\simeq$ 0.03) for the MB profile (the $\beta$) profile when 
the size of the halo is about 5 arcmin, just around the beam size of Planck. 
The signal to noise ratio will be $\sim$ 1 and 2 for the MB profile and $\beta$ 
profile, respectively.

\section{Conclusion and discussion}
\label{sec:discussion}
In this work, we investigate the kSZ effect of the CGM for MW-size halos. By 
weighting the CMB sky with the reconstructed velocity at the
galaxy position, one can pick out the kSZ effect of the CGM. The cross 
correlation is however subject to heavy contamination mainly from the primary 
CMB. To remove the large scale contaminations from the primary CMB, one can 
take advantage of different scale dependences between the primary CMB and the 
kSZ signal, and choose a minimum variance estimator by designing a matched 
filter. One can then stack the filtered map to further suppress the statistical 
error. With this method, we alternatively provide a statistical means to
detect the CGM via the kSZ effect. Furthermore, in its own right, kSZ effect of 
the missing galactic baryons can be cleanly measured, given field galaxies with 
similar host halo mass. Since current detection of the kSZ effect from the 
missing galactic baryons is mixed with the kSZ effect from ICM/IGM, a clean 
sample of galaxies, such as field galaxies from SKA, should be essential.

For comparison, we take the compelling MB profile as well as the empirical
$\beta$ model to test the detectability. The matched filter is designed
to remove the primary CMB on large scales, and keep the small scale kSZ signal
conserved. For the MB profile, given an ACT-like SZ survey and 10000 
Milky Way-size halos in a narrow redshift bin, we can achieve more than $1
\sigma$ for the CGM of MW-size halos at $z\sim0.5$. For galaxies occupying
a single pixel, the matched filter degenerates to the aperture photometry (AP)
filter, which can be easily implemented in the CMB map by subtracting the
average of an outer ring of equal area from the central part \citep[see][for an
example]{PlanckIntResCV}. At lower redshifts, more resolved pixels of each 
galaxy are involved to regulate the filter, the S/N increase greatly with the 
same mount of galaxies, and reaches a maximum S/N about 7.5 at $z\sim0.1$. The 
CGM with the $\beta$ model is more concentrating and the S/N turns out to be 
much higher. Though at high redshift the two profiles show similar S/N, the kSZ 
signal of $\beta$ profile has much higher S/N at $z<0.3$.
It can reach up to a maximum 19 $\sigma$ detection at a lower redshift (z$\sim$ 
0.03). 

Although it is unclear a steeper profile like the $\beta$ model is favoured by 
observational constraints for the CGM in Milky Way-size halos, kSZ effect of 
such profile can give insight to its upper limit. Due to the different 
responses of the minimum variance estimator to the filter, mainly at low 
redshift ($z<0.1$), it is possible to discriminate MB profile from the $\beta$ 
profile, as the underlying CGM will optimally respond to the corresponding 
filter profile and show the highest significance of detection among a set of 
filters. We also test a Planck-like survey, and with the same number of 
galaxies, it can only marginally detect the kSZ signal of the CGM at very low 
redshift. Though this work targets the CGM of late type galaxies, it can 
also serve to probe the kSZ effect for early type galaxies given the knowledge 
of the CGM's profile.

In this work, we have made a set of assumptions, and some of them are 
still under debate. We discuss several major concerns here.

\begin{enumerate}
 \item 
 {\it It is still under debate the total mass fraction and the spatial 
distribution of the CGM.} Such information depends on the halo mass, the 
accretion history, the star formation, the feedback energy deposited and the 
merging process. On one hand, simulations of galaxy formation use a broad 
baryon fraction ranging from 30-100\%. On the other hand, there are 
controversial constraints on baryon fraction $f_b$ within the virial radius. For 
example, \citet{Humphrey2011, Humphrey2012} found a consistent $f_b$ with the 
cosmic mean, while \citet{Anderson2011,Anderson2014} draw conclusions that only
~10-20\% missing baryons can be accounted for within the virial radius of the
host halo. The radial extent of X-ray observations, the spectral modeling, and
the metallicity distribution suggests different conclusions. Therefore, in 
this work, we take $f_{\rm CGM}=$62.5\%  of the cosmic mean, which is the missing 
baryon fraction of the Milky Way, as a fiducial fraction of the hot coronae 
constituting the missing part of baryons. We can easily scale the signal to 
noise ratio with $f_{\rm gas}$ with other values using Eq. 
\ref{eq:optical_depth} and Eq.  \ref{eq:signal_to_noise_filtered}.

 \item 
 {\it The uncertainty of the optical depth profile.} We can see that for a 
given SZ survey with fixed resolution, the two profiles of CGM lead to similar 
filter profile and thus similar S/N at high redshifts ($z\gtrsim 0.3$ for 
ACT, $z\gtrsim0.1$ for Planck) when galaxies are unresolved. When galaxies
become well resolved at lower redshifts, the more concentrating $\beta$ profile
has steeper kSZ effect, and the filter profiles diverge significantly. Since
either the MB profile or the $\beta$ model may not account for the true
profile of the underlying CGM, or the profiles of halos with the same
mass can vary due to different galaxy formation histories, the uncertainty in
the profile can induce variance in the estimator and degrade the
detectability. Future application of this method requires more observational
constraints on the average profile and the scatter of the profile for the CGM.

 \item
 {\it Mass range of host halos.} In this work, we have shown that for MW-size
 halos, the method works well. In real survey, host halos of galaxies 
 span a much broader mass range. This technique can be carried out with careful
 design.  On one hand, as shown in Eq. \ref{eq:optical_depth} that
 there's degeneracy between $f_{\rm gas}$ and M$^{2/3}$, high mass with low
 $f_{\rm gas}$ (e.g. beta profile for massive ellipticals) can lead to similar S/N
 for halos with low mass and high $f_{\rm gas}$. On the other hand, we can
 always bin host halos of a broad mass range into subsamples of narrow mass
 bin. We can then apply the filter to galaxies to obtain a measure of $\langle
 v_\parallel U \rangle$ for each subsample, and stack them to reduce the
 statistical noise further.

 \item 
 {\it The projection effect.} As a temperature distortion upon the CMB, 
the kSZ signal is an effect that involves projecting all the free electrons 
thus the ionized gas along the line of sight. Therefore, the kSZ effect of the
CGM would suffer from the contamination of the chance alignment of background
galaxies, clusters or diffuse baryons around galaxies, which are within the
velocity correlation length \citep[see][for the observational evidence of
baryons beyond the virial radius of halos]{PlanckIntResCV}. One can in
principle model the contribution these sources and filter them out, by
accounting for the two-halo term of the kSZ effect, given the cosmology
and the halo occupation distribution for late type galaxies or the distribution
of the ICM for clusters. For example, \citet*{Singh2015} calculated the power
spectrum of the kSZ from the CGM and clusters using halo model, and they found
the kSZ signal from CGM would only dominate the clusters on scales $l>10000$.
We can alternatively compute the cross correlation between the ICM and the CGM
using hydrodynamic simulations, and quantify the influence of the projection of
ICM along the line of sight of the CGM. Another possible contamination is the
potentially prevailing IGM or filamentary structures bridging galaxies beyond
the virial radius. For field galaxies from SKA HI survey, this problem is
less significant. The question is still open whether the missing baryons 
in galaxies are detained in the CGM or have been expelled to the IGM, or have 
never been accreted into galaxies in the galaxy formation process in the first 
place. The projection effect is thus to some extent degenerate to the gas 
fraction $f_{\rm gas}$ within the virial radius. Taking $f_{\rm gas}$ as a free 
parameter and assigning $f_{\rm gas}$ with some empirical distribution would be
a reasonable choice.

  \item 
  {\it The stochasticity in the velocity correlation.} A number of issues can 
induce uncertainties in the velocity correlation. We test the velocity 
reconstruction and the cross correlation under the influence of the redshift 
space distortion, the shot noise, and the gas physics using a set of
hydrodynamic simulations. We have found these effect will not fail the
reconstruction. We have also assumed a deterministic galaxy bias $b_g$ in the
velocity reconstruction. Though there are stochasticities in $b_g$, the
uncertainties are under control \citep{Bonoli2009} on scales of interest. As 
discussed in \S \ref{subsec:velocity_dispersion}, the peak contribution to the 
velocity correlation is from around $k\sim 0.05$\hMpc, and 70\% contribution is 
from $k\lesssim 0.1$\hMpc. Given a galaxy survey with number density larger than
$10^{-3} h^3\,$Mpc$^{-3}$, the recovered velocity traces the underlying 
velocity very well on scales of interest, as shown in Fig. \ref{fig:coef}, and
we take $r_{uv}=0.7$ as a fiducial value in the S/N estimation. The application
of the velocity reconstruction to the real survey has also proved effective
\citep{PlanckIntResCV}.

 \item 
 {\it The number density and the redshift distribution of galaxy surveys}.  
The realistic measurement of the kSZ signal of CGM of galaxies will 
depend on the number density and the redshift distribution of the galaxies of
the real survey. On one hand for example, SKA \citep{Meyer2015SKA, Abdalla2015}
will provide a wide and deep survey of HI galaxies covering the ACT/SPT sky
partly, and map $10^8$ field galaxies. It will be sufficient to recover the 
velocity field from the density field, and thus recover the kSZ effect of the 
CGM. On the other hand, 2MASS redshift survey (2MRS) \citep{Erdogdu2006, 
Huchra2012} have independently completed the peculiar velocity survey, and 
there are roughly 6000 edge-on spiral galaxies in local universe. This 
catalogue alternatively provide a measure of the velocity field. However, the 
spiral galaxy number ($\sim$ 150) covering the ACT sky (780 deg$^2$ released in 
the southern sky) is so small that the CGM of these galaxies may not be 
detectable. With even smaller beam size and data release of full 2500 $\deg^2$ 
sky in the future, SPT can serve to probe the CGM of spiral galaxies in 2MRS.

 \item 
 {\it The CGM of early type galaxies.} The method presented in this paper can 
in principle apply to the CGM of all galaxies with given profiles. For example, 
the CGM in massive early type galaxies, even with low baryon fraction, can
contain more hot gas than in late type galaxies \citep{Anderson2013}. If this is
the prevailing case, early type galaxies should contribute a huge reservoir of
CGM for the kSZ effect. 6DF Galaxy Redshift Survey (6DFGRS) \citep{Springob2014,
Campbell2014} has measured the peculiar velocity of $\sim$ 9000 early type
galaxies at $z<0.055$ in the southern sky. A very rough estimate shows that
there are $\sim$ 300 galaxies in the area covering the ACT sky. Given the
$\beta$ profile, as a usually adopted profile for the CGM in elliptical
galaxies, the CGM in the halo with mass of $10^{12}$ M$_\odot$ (as shown in Fig.
\ref{fig:snz}) can be detected with $\sim 2$ sigma level with a rough estimate
via this stacking scheme. Massive Elliptical galaxies with higher mass can show
even higher S/N.

\end{enumerate}

\section*{Acknowledgement}
We thank Pengjie Zhang for helpful discussion, and Weipeng Lin for kindly
providing the simulation data. JS is supported by the National Natural Science
Foundation of China under grant No. 11203053. TF was partially supported by the 
National Natural Science Foundation of China grant No. 11273021, by the 
Strategic Priority Research Program ``The Emergence of Cosmological Structures''
of the Chinese Academy of Sciences, grant No. XDB09000000, and by the National 
Natural Science Funds for Distinguished Young Scholar grant No. 11525312.


\begin{thebibliography}{100}
\expandafter\ifx\csname natexlab\endcsname\relax\def\natexlab#1{#1}\fi

\bibitem[Abdalla et al.(2015)]{Abdalla2015}
Abdalla, F.~B., Bull, P., Camera, S., et al.\ 2015, Advancing Astrophysics with the Square 
Kilometre Array (AASKA14), 17

\bibitem[{Anderson \& Bregman(2011)}]{Anderson2011}
Anderson M.~E., Bregman J.~N., 2011, ApJ 2011, 737, 22

\bibitem[{{Anderson}, {Bregman} \& {Dai}(2013){Anderson}, {Bregman}, \&
  {Dai}}]{Anderson2013}
{Anderson} M.~E., {Bregman} J.~N., {Dai} X., 2013, \apj, 762, 106

\bibitem[Anderson \& Bregman(2014)]{Anderson2014} 
Anderson, M.~E., Bregman, J.~N.\ 2014, \apj, 785, 67


\bibitem[{{Baldauf} {et~al}\mbox{.}(2010){Baldauf}, {Smith}, {Seljak}, \&
  {Mandelbaum}}]{Baldauf2010}
{Baldauf} T., {Smith} R.~E., {Seljak} U., {Mandelbaum} R., 2010, \prd, 81,
  063531

\bibitem[{{Bernardeau}, {Pitrou} \& {Uzan}(2011){Bernardeau}, {Pitrou}, \&
  {Uzan}}]{Bernardeau2011}
{Bernardeau} F., {Pitrou} C., {Uzan} J.~P., 2011, \jcap, 2, 15

\bibitem[Bhattacharya \& Kosowsky(2008)]{Bhattacharya2008} 
Bhattacharya, S.,  Kosowsky, A.\ 2008, \prd, 77, 083004 

\bibitem[{Birkinshaw(1999)}]{Birkinshaw1999}
Birkinshaw M., 1999, Phys.Rept., 310, 97

\bibitem[{{Bonoli} \& {Pen}(2009)}]{Bonoli2009}
{Bonoli} S., {Pen} U.~L., 2009, \mnras, 396, 1610

\bibitem[{{Bregman}(2007)}]{Bregman2007}
{Bregman} J.~N., 2007, \araa, 45, 221

\bibitem[Bregman \& Lloyd-Davies(2007)]{Bregman2007b}
Bregman, J.~N.,  Lloyd-Davies, E.~J.\ 2007, \apj, 669, 990 

\bibitem[{{Buote} {et~al}\mbox{.}(2009){Buote}, {Zappacosta}, {Fang},
  {Humphrey}, {Gastaldello}, \& {Tagliaferri}}]{Buote2009}
{Buote} D.~A., {Zappacosta} L., {Fang} T., {Humphrey} P.~J., {Gastaldello} F.,
  {Tagliaferri} G., 2009, \apj, 695, 1351

\bibitem[{{Campbell} {et~al}\mbox{.}(2014){Campbell}, {Lucey}, {Colless},
  {Jones}, {Springob}, {Magoulas}, {Proctor}, {Mould}, {Read}, {Brough},
  {Jarrett}, {Merson}, {Lah}, {Beutler}, {Cluver}, \& {Parker}}]{Campbell2014}
{Campbell} L.~A. {et~al.}, 2014, \mnras, 443, 1231

\bibitem[{{Carlstrom}, {Holder} \& {Reese}(2002){Carlstrom}, {Holder}, \&
  {Reese}}]{Carlstrom2002}
{Carlstrom} J.~E., {Holder} G.~P., {Reese} E.~D., 2002, \araa, 40, 643


\bibitem[Cavaliere \& Fusco-Femiano(1976)]{Cavaliere1976} 
Cavaliere, A., Fusco-Femiano, R.\ 1976, \aap, 49, 137

\bibitem[{{Cen} \& {Ostriker}(1999)}]{CO1999}
{Cen} R., {Ostriker} J.~P., 1999, \apj, 514, 1

\bibitem[Cen et al.(2005)]{Cen2005} 
	Cen, R., Nagamine, K.,  Ostriker, J.~P.\ 2005, \apj, 635, 86

\bibitem[{Crawford {et~al}\mbox{.}(2014)Crawford, Schaffer, Bhattacharya, Aird,
  Benson, Bleem, Carlstrom, Chang, Cho, Crites, de~Haan, Dobbs, Dudley, George,
  Halverson, Holder, Holzapfel, Hoover, Hou, Hrubes, Keisler, Knox, Lee,
  Leitch, Lueker, Luong-Van, McMahon, Mehl, Meyer, Millea, Mocanu, Mohr,
  Montroy, Padin, Plagge, Pryke, Reichardt, Ruhl, Sayre, Shaw, Shirokoff,
  Spieler, Staniszewski, Stark, Story, van Engelen, Vanderlinde, Vieira,
  Williamson, \& Zahn}]{Crawford2013}
Crawford T.~M. {et~al.}, 2014, ApJ, 784, 143

\bibitem[Dai et al.(2012)]{Dai2012} 
	Dai, X., Anderson, M.~E., Bregman, J.~N.,  Miller, J.~M.\ 2012, \apj, 
755, 107


\bibitem[{{Danforth} \& {Shull}(2008)}]{Danforth2008}
{Danforth} C.~W., {Shull} J.~M., 2008, \apj, 679, 194

\bibitem[Dav{\'e} et al.(2001)]{Dave2001}
	Dav{\'e}, R., Cen, R., Ostriker, J.~P., et al.\ 2001, \apj, 552, 473

\bibitem[{{Erdo{\v g}du} {et~al}\mbox{.}(2006){Erdo{\v g}du}, {Lahav},
  {Huchra}, {Colless}, {Cutri}, {Falco}, {George}, {Jarrett}, {Jones}, {Macri},
  {Mader}, {Martimbeau}, {Pahre}, {Parker}, {Rassat}, \&
  {Saunders}}]{Erdogdu2006}
{Erdo{\v g}du} P. {et~al.}, 2006, \mnras, 373, 45

\bibitem[Faerman et al.(2016)]{Faerman2016} 
Faerman, Y., Sternberg, A., McKee, C.~F.\ 2016, arXiv:1602.00689 

\bibitem[{{Fang}, {Bullock} \& {Boylan-Kolchin}(2013){Fang}, {Bullock}, \&
  {Boylan-Kolchin}}]{Fang2013}
{Fang} T., {Bullock} J., {Boylan-Kolchin} M., 2013, \apj, 762, 20

\bibitem[{{Fang} {et~al}\mbox{.}(2010){Fang}, {Buote}, {Humphrey}, {Canizares},
  {Zappacosta}, {Maiolino}, {Tagliaferri}, \& {Gastaldello}}]{Fang2010}
{Fang} T., {Buote} D.~A., {Humphrey} P.~J., {Canizares} C.~R., {Zappacosta} L.,
  {Maiolino} R., {Tagliaferri} G., {Gastaldello} F., 2010, \apj, 714, 1715

\bibitem[{{Fang} {et~al}\mbox{.}(2002){Fang}, {Marshall}, {Lee}, {Davis}, \&
  {Canizares}}]{Fang2002}
{Fang} T., {Marshall} H.~L., {Lee} J.~C., {Davis} D.~S., {Canizares} C.~R.,
  2002, \apjl, 572, L127


\bibitem[Faucher-Gigu{\`e}re et al.(2011)]{FG2011} 
Faucher-Gigu{\`e}re, C.-A., Kere{\v s}, D.,  Ma, C.-P.\ 2011, \mnras, 417, 2982


\bibitem[{{Feindt} {et~al}\mbox{.}(2013){Feindt}, {Kerschhaggl}, {Kowalski},
  {Aldering}, {Antilogus}, {Aragon}, {Bailey}, {Baltay}, {Bongard}, {Buton},
  {Canto}, {Cellier-Holzem}, {Childress}, {Chotard}, {Copin}, {Fakhouri},
  {Gangler}, {Guy}, {Kim}, {Nugent}, {Nordin}, {Paech}, {Pain}, {Pecontal},
  {Pereira}, {Perlmutter}, {Rabinowitz}, {Rigault}, {Runge}, {Saunders},
  {Scalzo}, {Smadja}, {Tao}, {Thomas}, {Weaver}, \& {Wu}}]{Feindt2013}
{Feindt} U. {et~al.}, 2013, \aap, 560, A90

\bibitem[{{Forman}, {Jones} \& {Tucker}(1985){Forman}, {Jones}, \&
  {Tucker}}]{Forman1985}
{Forman} W., {Jones} C., {Tucker} W., 1985, \apj, 293, 102

\bibitem[Fukugita et al.(1998)]{Fukugita1998}
	Fukugita, M., Hogan, C.~J.,  Peebles, P.~J.~E.\ 1998, \apj, 503, 518 

\bibitem[{Fukugita \& Peebles(2004)}]{Fukugita2004}
Fukugita M., Peebles P., 2004, \apj, 616, 643

\bibitem[{Fukugita \& Peebles(2006)}]{Fukugita2006}
Fukugita M., Peebles P., 2006, \apj, 639, 590

\bibitem[{{Gatto} {et~al}\mbox{.}(2013){Gatto}, {Fraternali}, {Read},
  {Marinacci}, {Lux}, \& {Walch}}]{Gatto2013}
{Gatto} A., {Fraternali} F., {Read} J.~I., {Marinacci} F., {Lux} H., {Walch}
  S., 2013, \mnras, 433, 2749

\bibitem[{{Gramann}(1998)}]{Gramann1998}
{Gramann} M., 1998, \apj, 493, 28

\bibitem[{{Grcevich} \& {Putman}(2009)}]{Grcevich2009}
{Grcevich} J., {Putman} M.~E., 2009, \apj, 696, 385

\bibitem[{{Guedes} {et~al}\mbox{.}(2011){Guedes}, {Callegari}, {Madau}, \&
  {Mayer}}]{Guedes2011}
{Guedes} J., {Callegari} S., {Madau} P., {Mayer} L., 2011, \apj, 742, 76

\bibitem[{{Gupta} {et~al}\mbox{.}(2012){Gupta}, {Mathur}, {Krongold},
  {Nicastro}, \& {Galeazzi}}]{Gupta2012}
{Gupta} A., {Mathur} S., {Krongold} Y., {Nicastro} F., {Galeazzi} M., 2012,
  \apjl, 756, L8

\bibitem[Haehnelt \& Tegmark(1996)]{Haehnelt1996}
  Haehnelt, M.~G.,  Tegmark, M.\ 1996, \mnras, 279, 545

\bibitem[{Hand {et~al}\mbox{.}(2012)Hand, Addison, Aubourg, Battaglia,
  Battistelli, Bizyaev, Bond, Brewington, Brinkmann, Brown, Das, Dawson,
  Devlin, Dunkley, Dunner, Eisenstein, Fowler, Gralla, Hajian, Halpern, Hilton,
  Hincks, Hlozek, Hughes, Infante, Irwin, Kosowsky, Lin, Malanushenko,
  Malanushenko, Marriage, Marsden, Menanteau, Moodley, Niemack, Nolta, Oravetz,
  Page, Palanque-Delabrouille, Pan, Reese, Schlegel, Schneider, Sehgal,
  Shelden, Sievers, Sifon, Simmons, Snedden, Spergel, Staggs, Swetz, Switzer,
  Trac, Weaver, Wollack, Yeche, \& Zunckel}]{Hand2012}
Hand N. {et~al.}, 2012, Phys. Rev. Lett. 109, 041101 (2012)

\bibitem[Henley \& Shelton(2013)]{Henley2013} 
Henley, D.~B., Shelton, R.~L.\ 2013, \apj, 773, 92 

\bibitem[Henley et al.(2015)]{Henley2015} 
Henley, D.~B., Shelton, R.~L., Kwak, K., Hill, A.~S.,  Mac Low, M.-M.\ 2015, 
\apj, 800, 102 

\bibitem[Hern{\'a}ndez-Monteagudo \& Ho(2009)]{Monteagudo2009} 
Hern{\'a}ndez-Monteagudo, C., Ho, S.\ 2009, \mnras, 398, 790


\bibitem[Hern{\'a}ndez-Monteagudo et al.(2015)]{Monteagudo2015} 
Hern{\'a}ndez-Monteagudo, C., Ma, Y.-Z., Kitaura, F.-S., et al.\ 2015, 
arXiv:1504.04011 

\bibitem[Hern{\'a}ndez-Monteagudo et al.(2006)]{Monteagudo2006} 
Hern{\'a}ndez-Monteagudo, C., Verde, L., Jimenez, R., Spergel, D.~N.\ 2006, 
\apj, 643, 598 


\bibitem[Ho et al.(2009)]{Ho2009}
	Ho, S., Dedeo, S.,  Spergel, D.\ 2009, arXiv:0903.2845

\bibitem[{Hu {et~al}\mbox{.}(1998)Hu, Seljak, White, \&
  Zaldarriaga}]{Zaldarriaga1997}
Hu W., Seljak U., White M., Zaldarriaga M., 1998, \prd, 57, 3290

\bibitem[{{Huchra} {et~al}\mbox{.}(2012){Huchra}, {Macri}, {Masters},
  {Jarrett}, {Berlind}, {Calkins}, {Crook}, {Cutri}, {Erdo{\v g}du}, {Falco},
  {George}, {Hutcheson}, {Lahav}, {Mader}, {Mink}, {Martimbeau}, {Schneider},
  {Skrutskie}, {Tokarz}, \& {Westover}}]{Huchra2012}
{Huchra} J.~P. {et~al.}, 2012, \apjs, 199, 26

\bibitem[Humphrey et al.(2011)]{Humphrey2011}
Humphrey, P.~J., Buote, D.~A., Canizares, C.~R., Fabian, A.~C.,  Miller, J.~M.\ 
2011, \apj, 729, 53 

\bibitem[{{Humphrey} {et~al}\mbox{.}(2012){Humphrey}, {Buote}, {O'Sullivan}, \&
  {Ponman}}]{Humphrey2012}
{Humphrey} P.~J., {Buote} D.~A., {O'Sullivan} E., {Ponman} T.~J., 2012, \apj,
  755, 166


\bibitem[{{Kashlinsky} {et~al}\mbox{.}(2010){Kashlinsky}, {Atrio-Barandela},
  {Ebeling}, {Edge}, \& {Kocevski}}]{Kashlinsky2010}
{Kashlinsky} A., {Atrio-Barandela} F., {Ebeling} H., {Edge} A., {Kocevski} D.,
  2010, \apjl, 712, L81

\bibitem[{{Kashlinsky} {et~al}\mbox{.}(2008){Kashlinsky}, {Atrio-Barandela},
  {Kocevski}, \& {Ebeling}}]{Kashlinsky2008}
{Kashlinsky} A., {Atrio-Barandela} F., {Kocevski} D., {Ebeling} H., 2008,
  \apjl, 686, L49

\bibitem[{{Keisler}(2009)}]{Keisler2009}
{Keisler} R., 2009, \apjl, 707, L42

\bibitem[{{Kere{\v s}} {et~al}\mbox{.}(2005){Kere{\v s}}, {Katz}, {Weinberg},
  \& {Dav{\'e}}}]{Keres2005}
{Kere{\v s}} D., {Katz} N., {Weinberg} D.~H., {Dav{\'e}} R., 2005, \mnras, 363,
  2

\bibitem[{{Knox}(1995)}]{Knox1995}
{Knox} L., 1995, \prd, 52, 4307

\bibitem[Komatsu et al.(2009)]{Komatsu2009}
	Komatsu, E., Dunkley, J., Nolta, M.~R., et al.\ 2009, \apjs, 180, 330

\bibitem[Larson et al.(2011)]{Larson2011}
	Larson, D., Dunkley, J., Hinshaw, G., et al.\ 2011, \apjs, 192, 16 

\bibitem[{{Lavaux}, {Afshordi} \& {Hudson}(2013){Lavaux}, {Afshordi}, \&
  {Hudson}}]{Lavaux2013}
{Lavaux} G., {Afshordi} N., {Hudson} M.~J., 2013, \mnras, 430, 1617

\bibitem[{{Lewis} \& {Bridle}(2002)}]{Lewis2002}
{Lewis} A., {Bridle} S., 2002, \prd, 66, 103511

\bibitem[{{Lewis}, {Challinor} \& {Lasenby}(2000){Lewis}, {Challinor}, \&
  {Lasenby}}]{Lewis2000CAMB}
{Lewis} A., {Challinor} A., {Lasenby} A., 2000, \apj, 538, 473

\bibitem[Li et al.(2014)]{Li2014}
	Li, M., Angulo, R.~E., White, S.~D.~M.,  Jasche, J.\ 2014, \mnras, 443, 
2311

\bibitem[Ma \& Zhao(2014)]{Ma2014}
Ma, Y.-Z.,  Zhao, G.-B.\ 2014, Physics Letters B, 735, 402 

\bibitem[{{Mak}, {Pierpaoli} \& {Osborne}(2011){Mak}, {Pierpaoli}, \&
  {Osborne}}]{Mak2011}
{Mak} D.~S.~Y., {Pierpaoli} E., {Osborne} S.~J., 2011, \apj, 736, 116

\bibitem[{{Maller} \& {Bullock}(2004)}]{MB2004}
{Maller} A.~H., {Bullock} J.~S., 2004, \mnras, 355, 694


\bibitem[McGaugh et al.(2010)]{McGaugh2010}
	McGaugh, S.~S., Schombert, J.~M., de Blok, W.~J.~G.,  Zagursky, M.~J.\ 
2010, \apjl, 708, L14

\bibitem[Meyer et al.(2015)]{Meyer2015SKA} 
	Meyer, M., Robotham, A., Obreschkow, D., et al.\ 2015, Advancing Astrophysics with the Square 
Kilometre Array (AASKA14), 131 


\bibitem[{{Mody} \& {Hajian}(2012)}]{Mody2012}
{Mody} K., {Hajian} A., 2012, \apj, 758, 4

\bibitem[{{Moster} {et~al}\mbox{.}(2011){Moster}, {Macciò}, {Somerville},
  {Naab}, \& {Cox}}]{Moster2011}
{Moster} B.~P., {Macciò} A.~V., {Somerville} R.~S., {Naab} T., {Cox} T.~J.,
  2011, \mnras, 415, 3750

\bibitem[{{Owen} \& {Warwick}(2009)}]{Owen2009}
{Owen} R.~A., {Warwick} R.~S., 2009, \mnras, 394, 1741

\bibitem[{{Planck Collaboration} {et~al}\mbox{.}(2011){Planck Collaboration},
  {Ade}, {Aghanim}, {Arnaud}, {Ashdown}, {Aumont}, {Baccigalupi}, {Balbi},
  {Banday}, {Barreiro}, {Bartlett}, {Battaner}, {Benabed}, {Benoît}, {Bernard},
  {Bersanelli}, {Bhatia}, {Blagrave}, {Bock}, {Bonaldi}, {Bonavera}, {Bond},
  {Borrill}, {Bouchet}, {Bucher}, {Burigana}, {Cabella}, {Cardoso}, {Catalano},
  {Cayón}, {Challinor}, {Chamballu}, {Chiang}, {Chiang}, {Christensen},
  {Clements}, {Colombi}, {Couchot}, {Coulais}, {Crill}, {Cuttaia}, {Danese},
  {Davies}, {Davis}, {de Bernardis}, {de Gasperis}, {de Rosa}, {de Zotti},
  {Delabrouille}, {Delouis}, {Désert}, {Dole}, {Donzelli}, {Doré}, {Dörl},
  {Douspis}, {Dupac}, {Efstathiou}, {Enßlin}, {Eriksen}, {Finelli}, {Forni},
  {Fosalba}, {Frailis}, {Franceschi}, {Galeotta}, {Ganga}, {Giard}, {Giardino},
  {Giraud-Héraud}, {González-Nuevo}, {Górski}, {Grain}, {Gratton}, {Gregorio},
  {Gruppuso}, {Hansen}, {Harrison}, {Helou}, {Henrot-Versillé}, {Herranz},
  {Hildebrandt}, {Hivon}, {Hobson}, {Holmes}, {Hovest}, {Hoyland},
  {Huffenberger}, {Jaffe}, {Jones}, {Juvela}, {Keihänen}, {Keskitalo},
  {Kisner}, {Kneissl}, {Knox}, {Kurki-Suonio}, {Lagache}, {Lamarre}, {Lasenby},
  {Laureijs}, {Lawrence}, {Leach}, {Leonardi}, {Leroy}, {Lilje},
  {Linden-Vørnle}, {Lockman}, {López-Caniego}, {Lubin}, {Macías-Pérez},
  {MacTavish}, {Maffei}, {Maino}, {Mandolesi}, {Mann}, {Maris}, {Martin},
  {Martínez-González}, {Masi}, {Matarrese}, {Matthai}, {Mazzotta},
  {Melchiorri}, {Mendes}, {Mennella}, {Mitra}, {Miville-Deschênes}, {Moneti},
  {Montier}, {Morgante}, {Mortlock}, {Munshi}, {Murphy}, {Naselsky}, {Natoli},
  {Netterfield}, {Nørgaard-Nielsen}, {Novikov}, {Novikov}, {O'Dwyer}, {Oliver},
  {Osborne}, {Pajot}, {Pasian}, {Patanchon}, {Perdereau}, {Perotto},
  {Perrotta}, {Piacentini}, {Piat}, {Pinheiro Gon{\c c}alves}, {Plaszczynski},
  {Pointecouteau}, {Polenta}, {Ponthieu}, {Poutanen}, {Prézeau}, {Prunet},
  {Puget}, {Rachen}, {Reach}, {Reinecke}, {Remazeilles}, {Renault},
  {Ricciardi}, {Riller}, {Ristorcelli}, {Rocha}, {Rosset}, {Rowan-Robinson},
  {Rubiño-Martín}, {Rusholme}, {Sandri}, {Santos}, {Savini}, {Scott},
  {Seiffert}, {Shellard}, {Smoot}, {Starck}, {Stivoli}, {Stolyarov}, {Stompor},
  {Sudiwala}, {Sunyaev}, {Sygnet}, {Tauber}, {Terenzi}, {Toffolatti}, {Tomasi},
  {Torre}, {Tristram}, {Tuovinen}, {Umana}, {Valenziano}, {Vielva}, {Villa},
  {Vittorio}, {Wade}, {Wandelt}, {White}, {Yvon}, {Zacchei}, \&
  {Zonca}}]{Ade2011}
{Planck Collaboration} {et~al.}, 2011, \aap, 536, A18



\bibitem[Planck Collaboration et al.(2014)]{Planck2014IntXIII} 
Planck Collaboration, Ade, P.~A.~R., Aghanim, N., et al.\ 2014, \aap, 561, A97 

\bibitem[Planck Collaboration et al.(2016)]{PlanckIntResCV} 
Planck Collaboration, Ade, P.~A.~R., Aghanim, N., et al.\ 2016, \aap, 586, A140 


\bibitem[Planck Collaboration et al.(2015)]{Planck2015XIII} 
Planck Collaboration, Ade, P.~A.~R., Aghanim, N., et al.\ 2015, arXiv:1502.01589

\bibitem[{{Putman}, {Saul} \& {Mets}(2011){Putman}, {Saul}, \&
  {Mets}}]{Putman2011}
{Putman} M.~E., {Saul} D.~R., {Mets} E., 2011, \mnras, 418, 1575

\bibitem[Rauch et al.(1997)]{Rauch1997} 
Rauch, M., Miralda-Escud{\'e}, J., Sargent, W.~L.~W., et al.\ 1997, \apj, 489, 
7 


\bibitem[{Reichardt {et~al}\mbox{.}(2012)Reichardt, Shaw, Zahn, Aird, Benson,
  Bleem, Carlstrom, Chang, Cho, Crawford, Crites, de~Haan, Dobbs, Dudley,
  George, Halverson, Holder, Holzapfel, Hoover, Hou, Hrubes, Joy, Keisler,
  Knox, Lee, Leitch, Lueker, Luong-Van, McMahon, Mehl, Meyer, Millea, Mohr,
  Montroy, Natoli, Padin, Plagge, Pryke, Ruhl, Schaffer, Shirokoff, Spieler,
  Staniszewski, Stark, Story, van Engelen, Vanderlinde, Vieira, \&
  Williamson}]{Reichardt2012}
Reichardt C.~L. {et~al.}, 2012, 2012 ApJ, 755, 70

\bibitem[{{Sarazin}(1986)}]{Sarazin1986}
{Sarazin} C.~L., 1986, Reviews of Modern Physics, 58, 1

\bibitem[Shao et al.(2011)]{Shao2011b}
	Shao, J., Zhang, P., Lin, W., Jing, Y.,  Pan, J.\ 2011, \mnras, 413, 
628 

\bibitem[Shull et al.(2012)]{Shull2012}
	Shull, J.~M., Smith, B.~D.,  Danforth, C.~W.\ 2012, \apj, 759, 23

\bibitem[Sievers et al.(2013)]{Sievers2013ACT}
	Sievers, J.~L., Hlozek, R.~A., Nolta, M.~R., et al.\ 2013, \jcap, 10, 060

\bibitem[Singh et al.(2015)]{Singh2015}
	Singh, P., Nath, B.~B., Majumdar, S.,  Silk, J.\ 2015, \mnras, 448, 2384

\bibitem[{{Sommer-Larsen}(2006)}]{Sommer-Larsen2006}
{Sommer-Larsen} J., 2006, \apjl, 644, L1

\bibitem[Spergel et al.(2003)]{Spergel2003}
	Spergel, D.~N., Verde, L., Peiris, H.~V., et al.\ 2003, \apjs, 148, 175


\bibitem[{{Spitzer}(1956)}]{Spitzer1956}
{Spitzer} J.~L., 1956, \apj, 124, 20


\bibitem[Springel(2005)]{Springel2005} 
Springel, V.\ 2005, \mnras, 364, 1105

\bibitem[{{Springob} {et~al}\mbox{.}(2014){Springob}, {Magoulas}, {Colless},
  {Mould}, {Erdo{\u g}du}, {Jones}, {Lucey}, {Campbell}, \&
  {Fluke}}]{Springob2014}
{Springob} C.~M. {et~al.}, 2014, \mnras, 445, 2677

\bibitem[Stocke et al.(2013)]{Stocke2013}
	Stocke, J.~T., Keeney, B.~A., Danforth, C.~W., et al.\ 2013, \apj, 763, 148 

\bibitem[{{Strauss} \& {Willick}(1995)}]{Strauss1995}
{Strauss} M.~A., {Willick} J.~A., 1995, \physrep, 261, 271

\bibitem[{{Sunyaev} \& {Zel'dovich}(1972)}]{SZ1972}
{Sunyaev} R.~A., {Zel'dovich} Y.~B., 1972, Comments on Astrophysics and Space
  Physics, 4, 173

\bibitem[{{Tegmark} \& {de Oliveira-Costa}(1998)}]{Tegmark1998}
{Tegmark} M., {de Oliveira-Costa} A., 1998, \apjl, 500, L83

\bibitem[{{Thom} {et~al}\mbox{.}(2012){Thom}, {Tumlinson}, {Werk}, {Prochaska},
  {Oppenheimer}, {Peeples}, {Tripp}, {Katz}, {O'Meara}, {Ford}, {Davé},
  {Sembach}, \& {Weinberg}}]{Thom2012}
{Thom} C. {et~al.}, 2012, \apjl, 758, L41

\bibitem[{Tripp {et~al}\mbox{.}(2004)Tripp, Bowen, Sembach, Jenkins, Savage, \&
  Richter}]{Tripp2004}
Tripp T.~M., Bowen D.~V., Sembach K.~R., Jenkins E.~B., Savage B.~D., Richter
  P., 2004, arXiv preprint astro-ph/0411151

\bibitem[{{Tripp}, {Savage} \& {Jenkins}(2000){Tripp}, {Savage}, \&
  {Jenkins}}]{Tripp2000}
{Tripp} T.~M., {Savage} B.~D., {Jenkins} E.~B., 2000, \apjl, 534, L1

\bibitem[{{Tripp} {et~al}\mbox{.}(2008){Tripp}, {Sembach}, {Bowen}, {Savage},
  {Jenkins}, {Lehner}, \& {Richter}}]{Tripp2008}
{Tripp} T.~M., {Sembach} K.~R., {Bowen} D.~V., {Savage} B.~D., {Jenkins} E.~B.,
  {Lehner} N., {Richter} P., 2008, \apjs, 177, 39

\bibitem[{{Tumlinson} {et~al}\mbox{.}(2011){Tumlinson}, {Werk}, {Thom},
  {Meiring}, {Prochaska}, {Tripp}, {O'Meara}, {Okrochkov}, \&
  {Sembach}}]{Tumlinson2011}
{Tumlinson} J. {et~al.}, 2011, \apj, 733, 111

\bibitem[Tytler et al.(2004)]{Tytler2004} 
Tytler, D., Kirkman, D., O'Meara, J.~M., et al.\ 2004, \apj, 617, 1 

\bibitem[{{Werk} {et~al}\mbox{.}(2014){Werk}, {Prochaska}, {Tumlinson},
  {Peeples}, {Tripp}, {Fox}, {Lehner}, {Thom}, {O'Meara}, {Ford}, {Bordoloi},
  {Katz}, {Tejos}, {Oppenheimer}, {Davé}, \& {Weinberg}}]{Werk2014}
{Werk} J.~K. {et~al.}, 2014, \apj, 792, 8

\bibitem[{{White} \& {Frenk}(1991)}]{WhiteFrenk1991}
{White} S.~D.~M., {Frenk} C.~S., 1991, \apj, 379, 52

\bibitem[{{White} \& {Rees}(1978)}]{White1978}
{White} S.~D.~M., {Rees} M.~J., 1978, \mnras, 183, 341

\bibitem[Zhang \& Johnson(2015)]{Zhang2015jcap}
Zhang, P., Johnson, M.~C.\ 2015, \jcap, 6, 046 

\bibitem[Zhang \& Stebbins(2011)]{Zhang2011} 
Zhang, P.,  Stebbins, A.\ 2011, Physical Review Letters, 107, 041301



\end{thebibliography}
\bibliographystyle{mn2e}

\end{document}